\documentclass[acmsmall]{acmart}
\AtBeginDocument{%
  \providecommand\BibTeX{{%
    \normalfont B\kern-0.5em{\scshape i\kern-0.25em b}\kern-0.8em\TeX}}}

\setcopyright{acmcopyright}
\copyrightyear{2018}
\acmYear{2018}
\acmDOI{10.1145/1122445.1122456}

\usepackage{graphicx}
\usepackage{booktabs}
\usepackage{multirow}
\usepackage{soul,color}
\usepackage{subcaption}

\begin{document}


\begin{flushleft}
      {\color{red} To cite: Shruti Phadke, Mattia Samory,  Tanushree Mitra. 2021. Characterizing Social Imaginaries and Self-Disclosures of Dissonance in Online Conspiracy Discussion Communities. Proc. ACM Hum.-Comput. Interact. Computer Supported Cooperative Work (CSCW' 21), (accepted Jul 2021).}
      \vspace{10pt}
\end{flushleft}

\title[Social Imaginaries and Self-Disclosures of Dissonance in Conspiracy Discussions]{Characterizing Social Imaginaries and Self-Disclosures of Dissonance in Online Conspiracy Discussion Communities}




\author{Shruti Phadke}
\email{phadke@uw.edu}
\affiliation{%
  \institution{University of Washington}
  \city{Seattle}
  \country{USA}
  }

\author{Mattia Samory}
\email{mattia.samory@gesis.org}
\affiliation{%
  \institution{GESIS}
  \city{Cologne}
  \country{Germany}
  }

\author{Tanushree Mitra}
\email{tmitra@uw.edu}
\affiliation{%
  \institution{University of Washington}
  \city{Seattle}
  \country{USA}
  }


\begin{abstract}
Online discussion platforms offer a forum to strengthen and propagate belief in misinformed conspiracy theories. Yet, they also offer avenues for conspiracy theorists to express their doubts and experiences of cognitive dissonance. Such expressions of dissonance may shed light on who abandons misguided beliefs and under which circumstances. This paper characterizes self-disclosures of dissonance about QAnon---a conspiracy theory initiated by a mysterious leader ``Q'' and popularized by their followers ``anons''---in conspiracy theory subreddits. To understand what dissonance and disbelief mean within conspiracy communities, we first characterize their social imaginaries---a broad understanding of how people collectively imagine their social existence. 
Focusing on 2K posts from two image boards, 4chan and 8chan, and 1.2 M comments and posts from 12 subreddits dedicated to QAnon, we adopt a mixed-methods approach to uncover the symbolic language representing the \textit{movement}, \textit{expectations}, \textit{practices}, \textit{heroes} and \textit{foes} of the QAnon community. 
We use these social imaginaries to  create a computational framework for distinguishing belief and dissonance from general discussion about QAnon, surfacing in the 1.2M comments. 
We investigate the dissonant comments to typify the dissonance expressed along QAnon social imaginaries. Further, analyzing user engagement with QAnon conspiracy subreddits, we find that self-disclosures of dissonance correlate with a significant decrease in user contributions and ultimately with their departure from the community. 
Our work offers a systematic framework for uncovering the dimensions and coded language related to QAnon social imaginaries and can serve as a toolbox for studying other conspiracy theories across different platforms. We also contribute a computational framework for identifying dissonance self-disclosures and measuring the changes in user engagement surrounding dissonance. Our work can provide insights into designing dissonance based interventions that can potentially dissuade conspiracists from online conspiracy discussion communities.

\end{abstract}



\maketitle

\section{Introduction}
Conspiracy ideation has widespread consequences ranging from inducing distrust and paranoia in individuals to threats to national security \cite{hofstadter2012paranoid,sunstein2009conspiracy}. Recent riots at the U.S Capitol by QAnon conspiracy believers show how conspiracy theorizing can lead to harmful collective action. Conspiracy theorists utilize online platforms to enhance and reaffirm their conspiratorial beliefs by discussions with peers \cite{mortimer2017understanding,phadke2021makes}. However, conspiracy theories are plagued by inconsistencies and fallacies \cite{wood2012dead} that could induce a state of cognitive dissonance---belief in contradictory ideas \cite{festinger1959cognitive}. 
Consider, for example, the following comment left by a QAnon conspiracy community member on Reddit: 
\begin{quote}
    \small I really want to believe. But lately I have been wondering whether Q is a psyop launched by the clowns. If the april showers fail to happen, I’m going to take some time off from Q and re-evaluate my interest.
\end{quote}

The above statement contains some key aspects relevant to belief and dissonance in the QAnon conspiracy theory. The user indicates their desire to believe in QAnon but at the same time expresses dissonance by doubting the legitimacy and efficacy of the QAnon leader---Q. Cognitive dissonance can motivate people to change their behaviors and attitudes \cite{festinger1962theory,mcgrath2017dealing}. Thus, experiencing dissonance with conspiracy belief can trigger individuals to depart from conspiratorial views. Hence, studying how conspiracists express dissonance with their beliefs is crucial towards understanding the pathways of recovery from conspiracy theories. But how do we identify expressions of dissonance in conspiratorial discussions? 
The cryptic, symbolic language, often unintelligible to outsiders of the community, but widely used for communicating within the community, poses significant challenges in interpreting online conspiracy discourse \cite{butter2020routledge,samory2018conspiracies}.
For example, several parts of the above comment are unintelligible to the outsiders of QAnon discussion communities. ``Q'', here refers to the leader of the QAnon community who posts prophetic message on image boards for his followers to decipher. ``Clowns'' are corrupt FBI and CIA agent declared as enemies of the QAnon movement. ``April showers'' refer to the promise made by Q to their followers about the arrests of corrupt politicians. 

Coded language is typical of conspiracy discussions and the primary mechanism for conspiracists to sustain their social imaginaries---collectively imagined realities by a group of similar minded people \cite{leone2019semiotics,taylor2004modern}. Understanding the social imaginaries of conspiracists can reveal the components of their collective existence, such as, their knowledge construction practices, their expectations from reality, their legends and characterization of the outside word \cite{butter2020routledge}. 
Thus identifying such social imagines will provide the means to understand conspiratorial expressions of belief and disbelief/dissonance.
In this paper, we focus on the QAnon conspiracy theory and first ask:

\vspace{5pt}

\noindent \textbf{RQ1: What are the social imaginaries of QAnon established by the leader Q?}

\vspace{5pt}
Given the prominence of Q in establishing the entire belief system of the QAnon community \cite{Howthree87online}, we conduct a qualitative content analysis of over 2000 Q-drops---posts made by Q on 4chan and 8chan---and lay out the social imaginaries that Q puts forth. We find five dimensions of QAnon social imaginaries: \textit{movement}, i.e., the collective identity of the believers who mobilize around the conspiracy; \textit{expectations}, i.e., promises and prophecies made by Q to their followers; \textit{practices}, i.e., collective knowledge construction and conspiracy theorizing of the QAnon community; \textit{heroes}, who are considered the leaders  serving the greater good in the social imaginaries; and \textit{foes} are the enemies of the QAnon community. 

Social imaginaries help conspiracists maintain separation between conspiracy theory insiders and outsiders \cite{leone2017fundamentalism}. Take, for instance, the use of the term ``clown'' in the previous example. Clown is connoted negatively so as to show inauthentic and unreliable behavior, thus distancing them as outsiders. 
On a meta-linguistic level, insider knowledge is essential for interpreting Q's posts and by extension to decode the expressions of other believers who follow Q. Thus, using the word ``clown'' that has a specific meaning inside the QAnon community, positions the author of the example as an insider of the QAnon community. In sum, social imaginaries help understand how QAnon believers frame their communication. Hence, we next ask:

\vspace{5pt}

\noindent \textbf{RQ2: How do QAnon followers communicate QAnon social imaginaries?}

\vspace{5pt}

To understand how QAnon followers adapt social imaginaries presented by their leader Q, we analyze Reddit communities where followers often reference and discuss Q-drops. 
Specifically, using the context of over 1.2M posts and comments from 12 QAnon discussion subreddits, we encode various phrases used to express the concepts of QAnon social imaginaries. We combine quantitative dynamic phrase matching techniques and manual validation to create the QAnon Canon---a lexicon of 403 phrases capturing coded language used by QAnon followers to communicate QAnon social imaginaries. For example, while in Q-drops, Hillary Clinton and Barack Obama, \emph{foes} of QAnon,  are mentioned as ``HRC'' and ``HUSSEIN'', QAnon followers adapt various expressions, such as ``Killary'', ``HC'' and ``Obummer'', ``fraudabama''. 

While symbolic communication of social imaginaries can facilitate secretive, in-group communication \cite{leone2019semiotics}, amplify the core conspiratorial ideas and beliefs towards them \cite{kimminich2016grounding}, they can also be used to express dissonance or disbelief in conspiracies \cite{leone2020semiotic}
For example, since Q is the leader of the QAnon conspiracy, doubt expressed by a QAnon follower questioning Q's legitimacy can be construed as dissonance with QAnon. 

\vspace{5pt}
\noindent \textbf{RQ3a: How can we identify belief and dissonance expressions in the QAnon community?}

\noindent \textbf{RQ3b: How do users express dissonance within the QAnon social imaginaries?}

\vspace{5pt}

Based on the QAnon Canon and other theoretically-informed constructs of belief and dissonance, such as the language of doubt or tentativeness \cite{evans2021expressions}, credibility cues \cite{mitra2017parsimonious}, integrative complexity \cite{czaja2016integrative},  we create a computational framework to classify Reddit comments as \textit{belief}, \textit{dissonance} or \textit{neutral}. Specifically, we use active learning techniques to surface and label expressions of belief and dissonance, and to distinguish them from general talk in QAnon subreddits. Our classifier achieved precision scores above 0.7 for all three classes. We find that phrases from QAnon Canon representing QAnon social imaginaries are important in predicting expressions of belief and dissonance.  For example, expressions of belief contain words related to QAnon \textit{movement} (``wwg1wga'' (where we go 1 we go all), ``patriots'') whereas dissonance self-disclosures frequently mention phrases related to \textit{expectations} (``arrests'', ``predictions'').
To further understand the fracture points in QAnon social imaginaries, we complement our quantitative method with a qualitative analysis of a sample of dissonant comments. Our analysis reveals
how dissonance occurs along the dimensions of QAnon social imaginaries. Specifically, we find that dissonance can be triggered because of unfulfilled \textit{expectations} and perceived illegitimacy of QAnon \textit{heroes}. 

Dissonance can also change behavior and attitudes \cite{festinger1959cognitive,festinger1962theory}. While some people strengthen their beliefs or even seek validation by recruiting more believers, others may choose to leave the community \cite{festinger1959cognitive}. Hence, we finally pose RQ4, where we analyze how user engagement inside and outside the QAnon communities change after self-disclosure of dissonance.  

\vspace{5pt}
\noindent \textbf{RQ4: How does engagement in QAnon subreddits change after expressing dissonance?}
\vspace{5pt}

We conduct an interrupted time series (ITS) analysis of user contributions---number of comments or posts---inside the QAnon communities in the 12 week period surrounding self-disclosures of dissonance. We find that user contributions in QAnon communities decrease significantly, immediately after dissonance disclosure, but not after expressing belief, while their overall engagement on Reddit stays the same. We corroborate and extend these results through statistical models, showing that not only disclosures of dissonance are followed by a decrease in contributions, but also by the departure of the users from the community. In particular, users who disclose dissonance disproportionately more than belief, are those most likely to leave the community.

Below we outline the contributions and implications of our work:

\begin{itemize}
    \item We develop a systematic framework for uncovering social imaginaries of conspiracies and for finding various language correlates of social imaginaries in conspiracy discourse (Figure \ref{fig:rq1meth}). 
    \item We offer the QAnon Canon, a lexicon of over 403 phrases capturing symbolic language and its shared meanings across QAnon social imaginaries that can serve as a toolbox for researchers to extend our study to other platforms (Table \ref{tab:canon_example}). 
    \item We offer a computational framework for identifying expressions of dissonance in the QAnon community (Figure \ref{fig:RQ2_method}). 
    \item We detail the points of fracture in QAnon social imaginaries (Table \ref{tab:rq3b_results}) which can be used to design dissonance based interventions for online conspiracy discussion participation. 
\end{itemize}




In the rest of the paper, we first provide background for the QAnon community and survey literature studying social imaginaries and dissonance in conspiracies. We then discuss the methods and results of each of the research questions and conclude with the discussion and ethical considerations. 


    

\section{Background}

\subsection{\textbf{What is QAnon?: Origin and Community Dynamics}}
In October 2017, a user on the image board, 4chan, signed off as ``Q'' and  posted a comment prophesying the arrests of Hillary Clinton and her staff.  In successive posts, ``Q'' purported the arrests of several other politicians associated with the ``Deep state''---a conspiracy theory claiming that a coalition of politicians in the U.S. run a shadow government involved in corruption and cronyism \cite{Blamingt53online}. In the next several months, the discussions around Q's posts took 4chan by the storm. Q presented themselves as an anonymous, high ranking U.S military official, with insider information about the U.S. government.  
Together, ``Q'', a mysterious, prophetic leader and ``anons'', Q's followers comprise the QAnon community. Specifically, Q predicts various political events in their Q-drops---messages posted to image boards such as 4chan, 8chan. Subsequently, followers of QAnon start piecing together the clues left in Q's posts and predictions. 
In fact, Q-drops are carefully crafted to contain cryptic messages such as ``\textit{the wormhole goes deep}'' or ``\textit{future proves past}'' which are meant to be clues for the followers to decipher. By encouraging followers to ``open their eyes'' and ``search for truth'' Q has institutionalized knowledge production practices that are unparalleled by other conspiracy movements \cite{partin2020construction}. Specifically, Q-followers are called ``bakers'' who assemble the ``crumbs'' (clues) left by Q into coherent pieces \cite{partin2020construction,amarasingam2020qanon}. This social construction of knowledge then produces unambiguous certainty through alternate reality \cite{partin2020construction}, a characteristic commonly associated with new religious movements \cite{barker1999new}. In this shared alternate reality, Q and their audience identify themselves as actors in a larger movement by using specific designations such as ``anons'', ``patriots'', ``bakers''.  A large part of the QAnon's alternate reality and worldview is represented by the symbolic language whose shared meaning is understood only within the community. 

QAnon community has started attracting research attention due to its clear mobilizing potential. Specifically, previous studies explored  the topics and dissemination of Q's messages on various online platforms and found that QAnon borrows theories from other conspiracies such as Pizzagate \cite{papasavva2020qoincidence}, shares moral values with Christian theology \cite{miller2021characterizing} and QAnon followers are likely to use violent rhetoric on Twitter \cite{planck2020we}. While the existing studies provide valuable characterization of the QAnon movement across platforms \cite{aliapoulios2021gospel}, deeper psychological and sociological exploration is required to deter increased engagement with QAnon \cite{garry2021qanon}. Our work fills this gap by first establishing QAnon social imaginaries symbolizing collective interpretation of reality by QAnon and then using those imaginaries to identify expressions of dissonance in the QAnon community. We start by first providing the background on the role of social imaginaries in the conspiracy communities.

\subsection{\textbf{Conspiracists and Social Imaginaries}}
Conspiracy theorizing generally consists of a belief that a covert operation is being carried out by a group of conspirators or secret organizations, to influence events \cite{keeley1999conspiracy,pigden1995popper}. Conspiracy theorizing is able to produce a certain aesthetic pleasure that enables people to form social imaginaries---coherent, collective imagination of social existence by a set of people \cite{leone2017fundamentalism,thompson1984studies}. 
Social imaginaries refer to the ways people imagine their social existence, their relationship between different social groups, deeper normative notions and the expectations of reality born out of such norms \cite{taylor2004modern}. 
For example, a social imaginary of a conspiracy theorist can consist of irrational interpretation of reality, such as, ``there is a global lobby trying to enslave common people'' or ``conspiracists being ridiculed by ignorant mass are further proof of the subversion by elites'' \cite{leone2017fundamentalism} or QAnon's purported worldview that ``America is run by a cabal of pedophiles and Satan-worshippers who run a global child sex-trafficking operation and QAnon are force of good stopping them'' \cite{WhatisQA22online}. 
In sum, social imaginaries lie at the heart of the belief systems and are a way for groups to rationalize their sense of reality and even find purpose in the collective action \cite{taylor2004modern}. Thus we argue that it is important to understand the conspiracists' social imaginaries. 

In this paper, we study conspiracy social imaginaries by focusing on the QAnon movement. The entire QAnon conspiracy theory is based on the leader Q's early messages posted on image boards such as 4chan and 8chan \cite{QAnonHas22online}. In fact, QAnon's social construction of knowledge is largely centered around understanding Q's missives \cite{zuckerman2019qanon}. Hence, in RQ1, to uncover the dimensions of QAnon social imaginaries, we perform a qualitative analysis of over 2000 posts made by Q on 4chan and 8chan. Social imaginaries are typically shared by a group of people and instill the sense of legitimacy to their cause and existence as a group \cite{taylor2004modern}. While in RQ1 we identify social imaginaries based on the leader Q's messages, Q's followers may adopt various linguistic expressions to communicate QAnon social imaginaries on online platforms. Hence, we next consider the role of semiotics---use of symbols to communicate shared meanings---in conspiracy online discourse.

\subsection{\textbf{Conspiracies and Semiotics}}
Maintaining the separation between the insiders---people who are aware of the conspiracy---and the outsiders---enemies of irrational interpretation of reality---is essential for sustaining and communicating the social imaginaries of conspiracy theories \cite{leone2017fundamentalism}. Semiotics, in the form of secretive communication and symbolic language, is used to maintain separation between the insiders and the outsiders of the conspiracy communities \cite{butter2020routledge,leone2017fundamentalism}. Specifically the type of semiotics based on paranoid thinking commonly associated with conspiracists \cite{hofstadter2012paranoid} is destined to render the communication unintelligible to the outsiders \cite{eco1979theory}. In fact, \citeauthor{leone2019semiotics} argues that the main role of semiotics in conspiracies is to disrupt the common sense interpretations of public discourse known to the outsiders. Consider for example the use of the words ``clowns'' or ``swamp'' in the QAnon community. While to the outsiders clowns mean comic performers and swamp means a wetland, in QAnon community ``clowns'' mean FBI or CIA agents and ``swamp'' refers to the collective of people and communities associated with deep state. By disrupting the common sense of the outside world, conspiracists create common sense of their own, shared among the community of insiders \cite{butter2020routledge,heathershaw2012national}. In other words, through the use of semiotics, conspiracists create, develop, and propagate shared meanings that sustain their social imaginaries \cite{butter2020routledge,leone2017fundamentalism}. 
In RQ2, we quantitatively identify semiotic patterns used by QAnon followers and encode them in the form of a phrase lexicon, QAnon Canon.


\subsection{Conspiracies and Cognitive Dissonance}
While social imaginaries and their semiotic communication play an important role in affirming conspiracy belief \cite{butter2020routledge}, they can be used to understand dissonance with conspiracies as well. 
While not all conspiracies are false or impossible \cite{basham2006malevolent}, many suffer inconsistencies, contradictions, and general epistemological challenges \cite{sunstein2009conspiracy,wood2012dead}. 
Realizing such inconsistencies may induce a state of dissonance among conspiracy followers.
How do conspiracy believers react when their beliefs are contradicted? Researchers found that mistrust in authorities or governing bodies is sufficient to overwhelm the contradictions between individual conspiracy theories \cite{wood2012dead}. 
Festinger coined the phenomenon of believing in contradictory, inconsistent ideas as ``cognitive dissonance'' \cite{festinger1962theory,festinger2017prophecy}. In a famous immersive ethnographic study, Festinger and colleagues infiltrated a UFO religion in Chicago where the cult leader had prophesied that the world will end on December 1954. Festinger and colleagues revealed that after the prophesied date and obvious signs of world not ending, members of the group experienced cognitive dissonance. According to the theory of cognitive dissonance proposed by \mbox{\citeauthor{festinger1962theory}}, dissonance can result from various individual or social factors. For example, dissonance can result from involuntary or voluntary exposure to information that directly contradicts previously held beliefs. Further, the simple act of having to choose between two contradictory ideas can also intensify the experience of dissonance. Moreover, dissonance can also result from the conflict between individually held and socially accepted beliefs. When confronted with such dilemmas, individuals may change their perception of contradictory ideas, find overlap between the two ideas or completely reverse their previously held beliefs \mbox{\cite{festinger1959cognitive}}. Indeed, after experiencing dissonance, different believers reacted in different ways. While some strengthened their convictions and even recruited newer members, others left the cult. In other words, being in the state of dissonance, where beliefs are challenged or contradicted, may lead people to change their behaviors or attitudes \cite{festinger1962theory,priolo2019three,stone2008practice}.

Similar to the cult leader in Festinger's study, Q---the leader of QAnon---has made several predictions and prophecies that never came true \cite{qdropsonline}. For example, amongst hundreds of other predictions, Q's very first prediction about Hillary Clinton's arrest in 2017 has provably failed \cite{qdropsonline}. Can Q's failed predictions, similar to the UFO religion studied by \mbox{\citeauthor{festinger1962theory}}, induce dissonance in Q's followers? What other fracture points in QAnon belief can induce cognitive dissonance?
In this light QAnon makes for an ideal case to study dissonance in conspiracies where we can analyze both, social imaginaries created by Q and the self-disclosure of dissonance by Q followers on the fracture points of the social imaginaries. To our knowledge, there is no study exploring how conspiracy believers express doubt, either in the face of contradictions or other ideals clashing with the social imaginaries. In RQ3, we fill this gap by first identifying the expressions of dissonance among the members of online QAnon community and next, highlighting the points of dissonance in the QAnon belief.

What happens after people express dissonance? Researchers studying addictive behaviors found that higher levels of cognitive dissonance can help people deconstruct their previously held norms and beliefs and pave the pathway for recovery\cite{bliuc2017building,galanter2014alcoholics,vaghefi2017addiction}. In RQ4, we analyze changes in user engagement with QAnon discussion communities after they express dissonance with QAnon.

\begin{figure*}[t]
    \centering
    \includegraphics[width=0.95\textwidth]{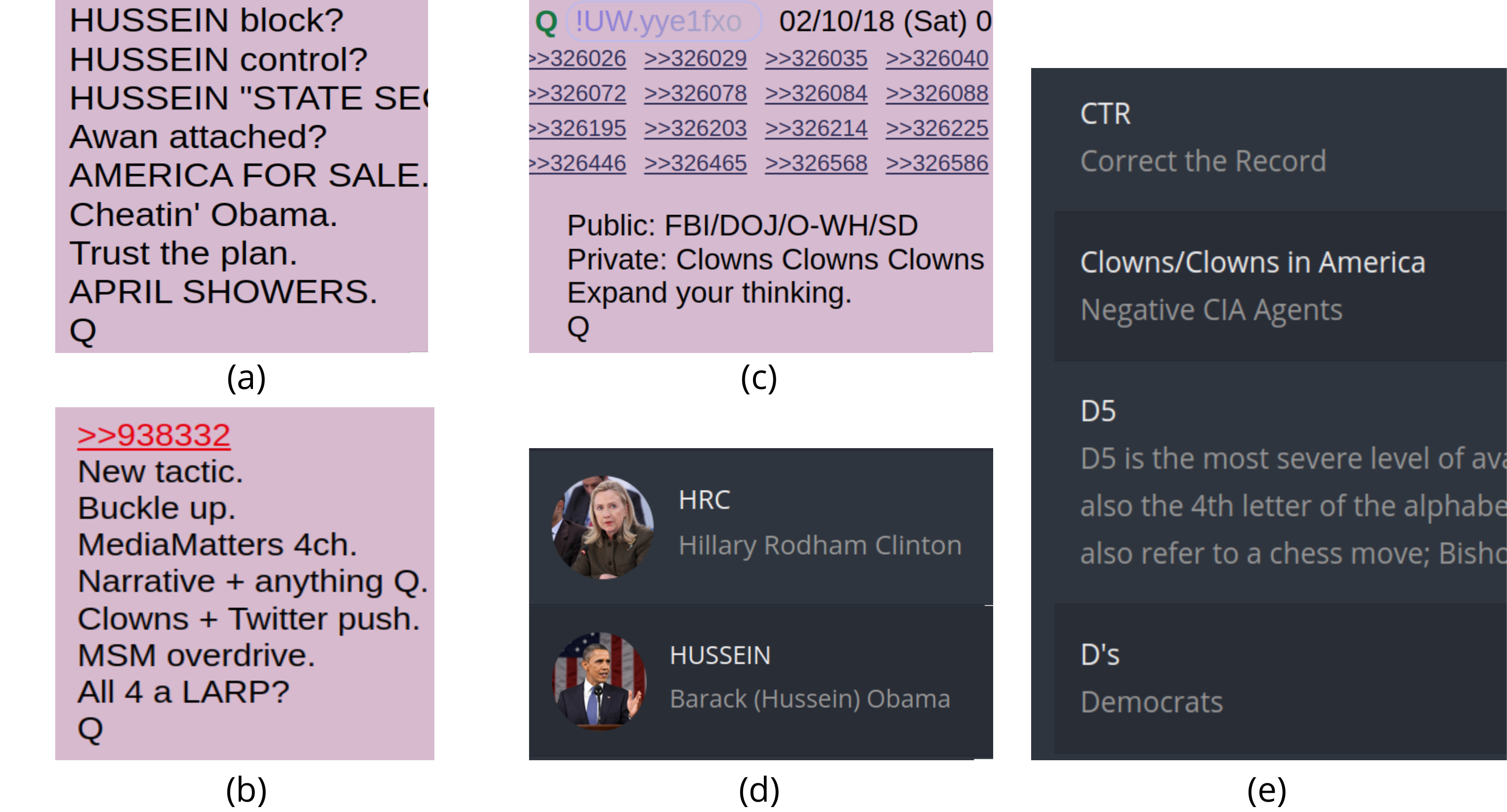}
    \caption{Example of Q-drops (a, b, c) and explanations of cryptic language (d, e) available on a QAnon aggregator site---{\small \tt qalerts.app}. In some cases, the meanings behind codes are explained in the Q drops itself. For example the word ``clowns'' mentioned in (b) is later explained in another Q-drop (c). 
    A common pursuit of the QAnon followers is to decode the cryptic language and understand the core message posted by Q. For example, on {\small \tt qalerts.app} one of the codewords ``HUSSEIN'' in a Q-drop (shown in (a)) is decoded by QAnon community members as Barack Hussein Obama (see (d)).}
    \label{fig:qdrops}
\end{figure*}

\section{Data}


\subsection{Q-drops Dataset: Q-drops from 4chan, 8chan}
In the beginning of the QAnon conspiracy, Q posted messages (Q-drops) on image boards, and the followers discussed Q-drops across various Reddit communities \cite{Howthree87online} (See Figure \ref{fig:qsubs}). 
In RQ1, we analyze Q-drops to characterize social imaginaries established by the leader Q for the QAnon community. Specifically, Q drops were made on 4chan and 8chan boards that are anonymous, ephemeral forums revolving around posting images along with text. 4chan and 8chan are infamous for hosting controversial content resulting in multiple temporary bans. We use {\tt qalerts.app}, a website that aggregates Q-drops from image boards. While the exact agency of {\tt qalerts.app} is unknown, it is a common resource used by QAnon communities \footnote{See QAnon and the Great Awakening group on Gab.com } and also by other researchers studying QAnon \cite{aliapoulios2021gospel}. There are several other Q-drop aggregation sites across internet however, most of the sites contain nearly similar record of Q posts (see Table 1 in \citeauthor{aliapoulios2021gospel} \cite{aliapoulios2021gospel}). 
We download first 2166 Q-drops that were made between October 2017 and September 2018 to allow for consistent analysis time period between various RQs in the paper \footnote{The last posted Q-drop was in December 2020, making it a total of 4953 Q-drops posted since 2017.}. For example, see subreddit timelines in Figure \ref{fig:qsubs}. The 12 QAnon discussions subreddits existed in the overall time period of November 2017 to  September 2018. Hence, we consider the first 2166 Q-drops that span over the same time period. 


{\tt qalerts.app} also hosts other resources, such as list of abbreviations specific to the QAnon community and research compiled by QAnon followers. See for example the screenshot of abbreviations provided on {\tt qalerts.app} in Figure \ref{fig:qdrops} (d) and (e). 
We utilize these resources to understand and contextualize the content of the Q-drops in RQ1. 

\subsection{QAnon Subreddits Dataset: Reddit Dataset of 12 Banned QAnon Communities}
 After the emergence of Q on 4chan in September 2017, the QAnon movement popularized on Reddit \cite{Howthree87online}. To reach a more mainstream audience, prominent QAnon followers created a subreddit called r/CBTS\_Stream, short for \textit{Calm before the Storm}--a popular saying in the QAnon community indicating impending arrests of the deep state politicians. However, r/CBTS\_Stream was banned in March 2018 for violating Reddit's terms of content policies. This ban resulted in the creation of new subreddits such as r/greatawakening2, r/BiblicalQ and others that combined accrued more subscribers than the original r/CBTS\_Stream \cite{Redditba46online}. Finally, 17 of these new communities were also banned by Reddit in September 2018 for repeated violation of content policies \cite{Redditba46online}.
 
\begin{figure*}[t]
    \centering
    \includegraphics[width=0.85\textwidth]{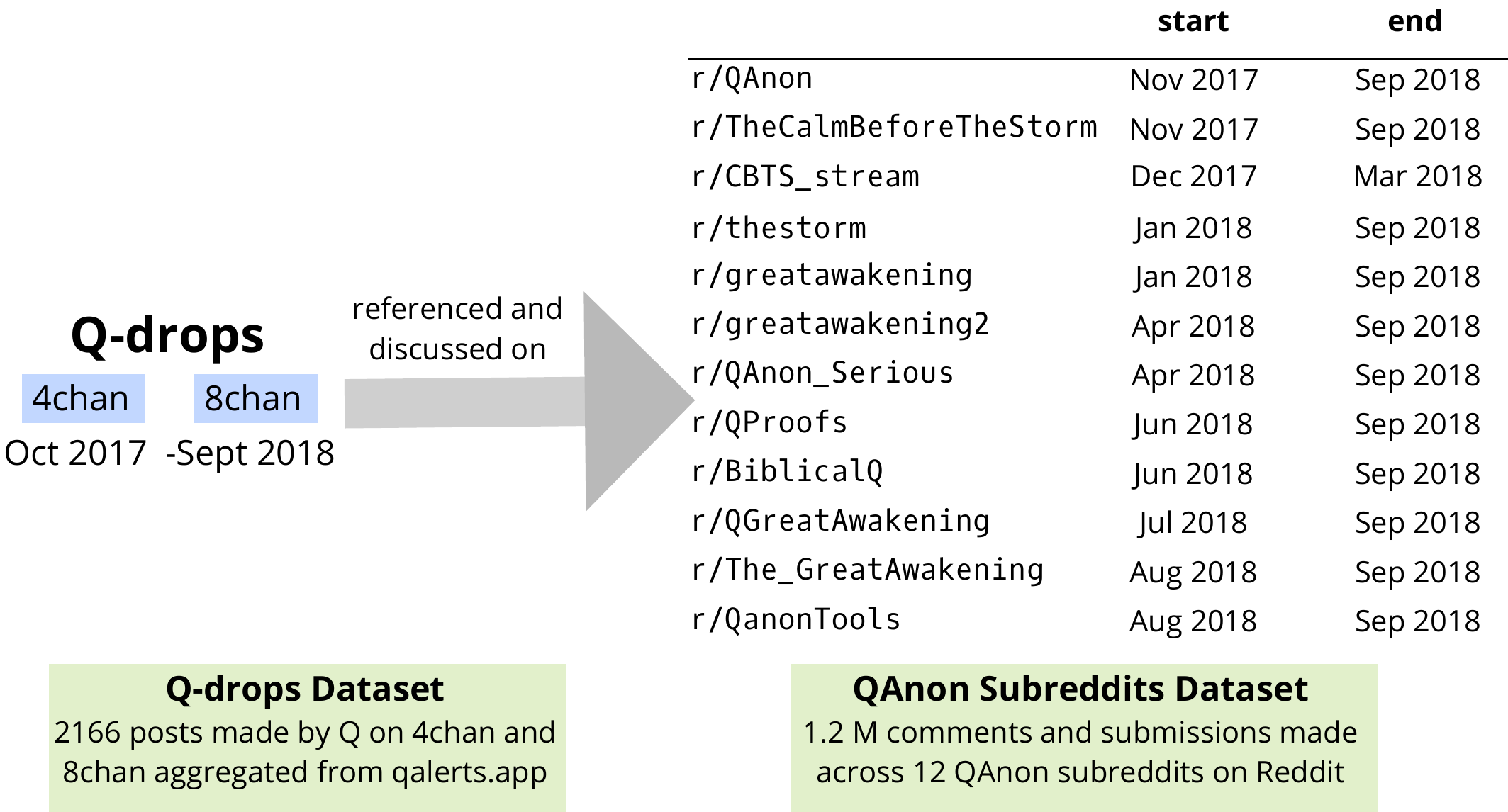}
    \caption{Figure shows two datasets used in this paper. The Q-drops dataset is used to derive social imaginaries in RQ1 while QAnon subreddit dataset is used in RQ2, RQ3 and RQ4. Usually, Q posts drops---Q-drops---on image boards such as 4chan and 8chan. Q-drops are then referenced and discussed in QAnon discussion subreddits. The 12 subreddits in our dataset comprise of posts made between November 2017 and September 2018. Hence, to understand relevant social imaginaries, we consider Q-drops from the start of Q in 2017 to September 2018.}
    \label{fig:qsubs}
\end{figure*}

 We identified the 17 banned QAnon related subreddits from various press mentions \mbox{\cite{Redditba46online,QAnonFol61online}}. To obtain the data for banned subreddits, we used Reddit Pushshift Dataset\footnote{\url{https://files.pushshift.io/reddit/comments/}} \cite{baumgartner2020pushshift}. Specifically, we queried the Pushshift data from Google Bigquery and obtained the submissions and comments from 12 of the 17 banned subreddits. The data for the rest of the 5 subreddits is not present on Pushshift nor through the official Reddit APIs. In total, we have 96,068 submissions and 1,104,096 comments made by 33,561 users across 12 subreddits listed in Figure \ref{fig:qsubs}. 


\section{RQ1: Characterizing the QAnon Social Imaginaries Established by Q}

\begin{figure*}[t]
    \centering
    \includegraphics[width=0.90\textwidth]{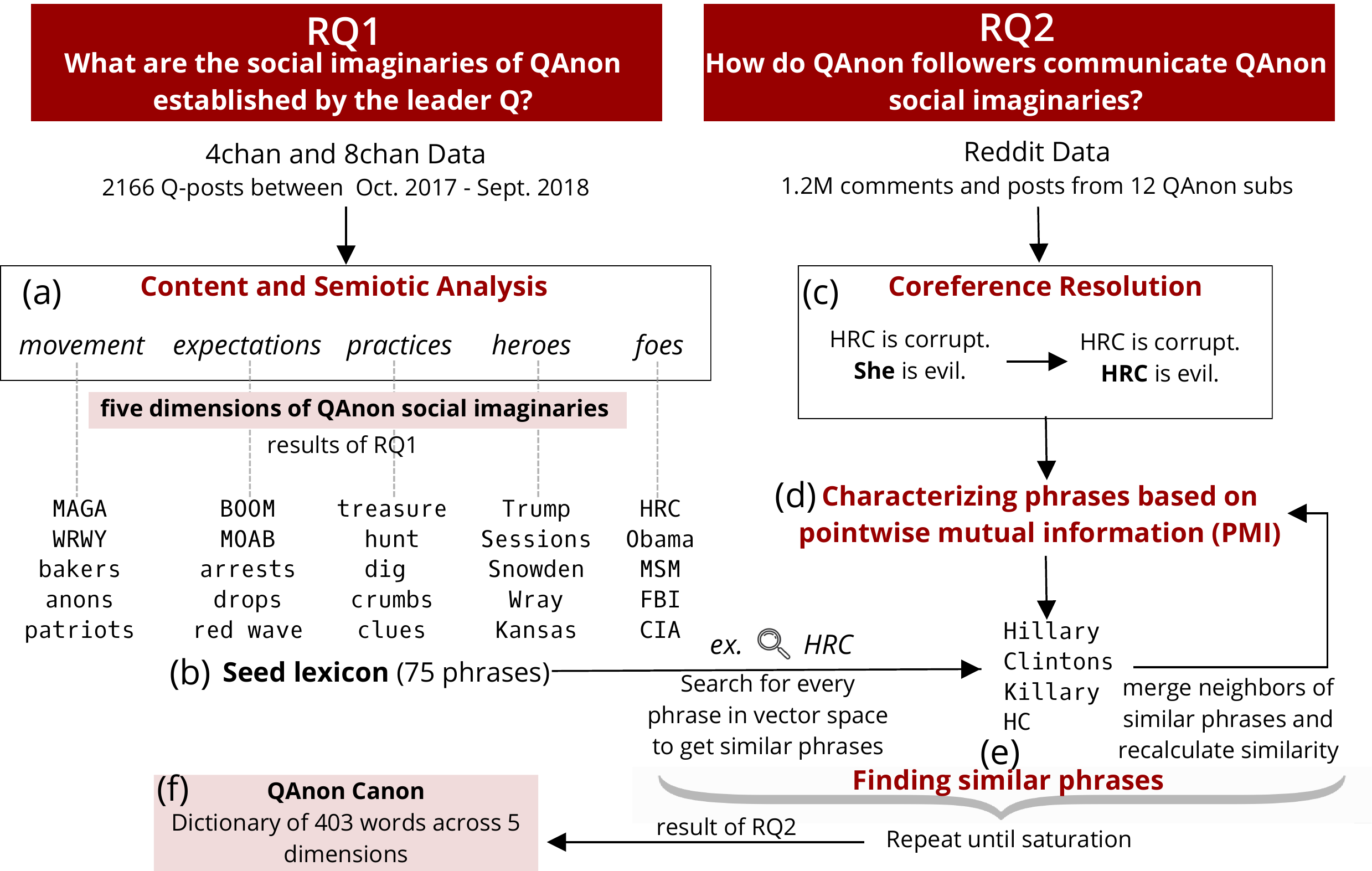}
    \caption{Figure showing our process for understanding the dimensions of QAnon social imaginaries and recording the ways in which social imaginaries are communicated by the followers inside the QAnon communities. (a) In RQ1 we perform content analysis of over 2K Q-drops and find five dimensions of the social imaginaries. (b) We also recorded 75 phrases---seed lexicon---that use coded language across various dimensions of social imaginaries. In RQ2, we expand the seed lexicon using public discourse in QAnon subreddits. We first (c) resolve coreferences and then (d) find numerical representations for meanings of phrases using PMI. (e) We then iteratively find similar phrases and (e) create QAnon Canon---a lexicon of 403 phrases capturing symbolic communication of QAnon social imaginaries}
    \label{fig:rq1meth}
\end{figure*}

What are the social imaginaries of the QAnon community? By understanding the social imaginaries of QAnon, we will uncover how QAnon perceive their community and the outside world, and how they construct their practices and form their expectations from the reality \cite{taylor2004modern}. We first describe our approach to understanding social imaginaries or worldview of the QAnon community. We employ qualitative content analysis to explore various dimensions of the QAnon social  imaginaries. Given the prominent use of semiotics in conspiracy discussions \cite{butter2020routledge,leone2017fundamentalism}, we focus on discovering words and phrases with shared meaning that would represent various aspects of the social imaginary. Finally, we present the five dimension of QAnon social imaginaries surfaced from the qualitative analysis along with examples. 



\subsection{RQ1 Method: Qualitative Content and Semiotic Analysis}
Q-drops---posts made on image boards by the leader Q---are at the heart of the QAnon movement. QAnon followers consider Q-drops as the main source of insider knowledge \cite{qdropsonline} and devote themselves to decoding Q-drops in order to expose the deep state \cite{aliapoulios2021gospel}. Hence, to understand the social imaginaries of the QAnon community, we analyze the Q-drops dataset (Figure \ref{fig:qsubs}). 
Specifically, we use inductive qualitative content analysis \cite{elo2008qualitative,weber1990basic}---an inductive reasoning method aimed at revealing themes, patterns and meanings embedded in the data \cite{lune2017qualitative}. The core idea behind content analysis is that many words or phrases from the text (Q-drops in this case) are categorized into much fewer coherent categories \cite{smithson2013gamechangers}. Words or phrases in the same category represent similar themes. This similarity can be based on semantic characteristics of the words and phrases \cite{kaplan1943content}. Further, to understand the process of semiosis within the QAnon community, we consider symbolic words and phrases that mediate a common meaning \cite{morris1938foundations}. Consider for example, the Q-drops in Figure \ref{fig:qdrops}. In (a), the phrase ``april showers'' may not mean anything to an outsider. However, inside the QAnon community ``april showers''  has specific significance; it symbolizes Q's prophecy about mass arrests of the deep state politicians in April 2018.  

\subsubsection{\textbf{Extracting information from Q-drops}} 
For each of the 2166 Q-drops, we first manually extracted words and phrases that contributed to shaping the narrative of the QAnon movement. For example, we recorded terms such as ``april showers'' \footnote{\url{https://qalerts.app/?n=1007}}, ``WWG1WGA'' \footnote{\url{https://qalerts.app/?n=1025}} (Where we go one we go all) and names of politicians associated with the deep state\footnote{\url{https://qalerts.app/?n=1708}}. 
After extracting words and phrases from the Q-drops we proceed to the iterative inductive coding phase. 

\subsubsection{\textbf{Iterative Inductive Coding}} After collecting words and phrases from Q-drops, we began to organize them into themes based on the semantic relationships. First two authors of the paper were involved in this inductive analysis. Our focus was on grouping words that represent similar meanings in the QAnon ideology.
For example, the words ``boom'', ``moab'' (mother of all booms), ``april showers'', ``red october'' all generally allude to a future event, significant to the take down of the deep state. Similarly, the words ``anons'', ``bakers'', ``autists'', ``patriots'' are all used to refer to insiders of the QAnon community. Note that, to understand the insider language of QAnon, we refer to the words cross-listed across various Q-drops and also other online resources such as; list of abbreviations on {\tt qalerts.app} and various news articles describing the Q-drops and QAnon language \cite{qdropsonline,QSpeakTh41online}. We iteratively developed the categories through a discursive process and found saturation at five categories that capture the social imaginaries of the QAnon community. Next, we explain the five dimensions of QAnon social imaginaries---\textit{movement}, \textit{expectations}, \textit{practices}, \textit{heroes}, and \textit{foes}---resulting from the qualitative analysis in detail along with the examples. 

\begin{table*}[t]
\centering
\resizebox{0.8\textwidth}{!}{%
\begin{tabular}{@{}ll@{}}
\toprule
category                      & example texts from Q-drops                                                                     \\ \midrule
\multirow{3}{*}{movement}     & \textit{Dear \textbf{Patriots}, We hear you. We hear all all Americans such as yourself}                \\ \cmidrule(l){2-2} 
                              & \textit{Think new arrivals. Proofs are important. Thank you, \textbf{autists} and \textbf{anons}}                \\ \cmidrule(l){2-2} 
                              & \textit{The  \textbf{wizards and warlocks} will not allow another Satanic Evil POS control our country} \\ \midrule
\multirow{3}{*}{expectations} & \textit{\textbf{HRC extradition} already in motion effective yesterday}                                 \\ \cmidrule(l){2-2} 
                              & \textit{Next week. \textbf{Boom. Boom. Boom.}}                                                          \\ \cmidrule(l){2-2} 
                              & \textit{Be vigilant today and expect a major \textbf{false flag}}                                       \\ \midrule
\multirow{3}{*}{practices}    & \textit{\textbf{Crumb} dump incoming fast. \textbf{Archive} immediately. Upload to graphic.}                     \\ \cmidrule(l){2-2} 
                              & \textit{Shall we play a game? Find @Snowden. Happy \textbf{Treasure Hunt}!}                             \\ \cmidrule(l){2-2} 
                              & \textit{Keep \textbf{digging} and keep \textbf{organizing} the info into graphics (critical).}                   \\ \midrule
\multirow{3}{*}{heroes}       & \textit{America First. This Is What Happens When  \textbf{POTUS} Has No Strings Attached.}              \\ \cmidrule(l){2-2} 
                              & \textit{Panic in DC. Trust  \textbf{Sessions}. Enjoy the show.}                                         \\ \cmidrule(l){2-2} 
                              & \textit{Trust \textbf{Sessions}. Trust \textbf{Wray}. Trust \textbf{Kansas}. Trust \textbf{Horowitz}. Trust \textbf{Huber}.}                \\ \midrule
\multirow{3}{*}{foes} & \textit{Realize \textbf{Soros}, \textbf{Clintons}, \textbf{Obama}, \textbf{Putin}, etc. are all controlled by 3 families} \\ \cmidrule(l){2-2} 
                              & \textit{Why does the \textbf{MSM} portray the country as being divided?}                                \\ \cmidrule(l){2-2} 
                              & \textit{Why wasn’t \textbf{HRC} prosecuted for the emails? These people are EVIL}                       \\ \bottomrule
\end{tabular}%
}
\caption{Five dimensions of QAnon social imaginaries along with the examples. These dimensions capture the collective identity of the QAnon community (\textit{movement}), QAnon's leaders and legends (\textit{heroes}), QAnon's enemies (\textit{foes}), QAnon knowledge construction practices (\textit{practices}) and the expectations from the reality (\textit{expectations})}
\label{tab:content_analysis_examples}
\end{table*}

\label{sec:soc_image_results}
\subsection{RQ1 Results: QAnon Social Imaginaries Established by the Leader Q}
We found five high level categories that represent the QAnon social imaginaries put forward by Q: \textit{movement}, \textit{expectations}, \textit{practices}, \textit{heroes}, and \textit{foes}. 
Table \ref{tab:content_analysis_examples} lists all dimensions of QAnon imaginaries and provides example Q-drops for each. 

\begin{enumerate}
    \item \textbf{Movement:}  The \textit{movement} category signifies the collective identity of the QAnon community. Movement includes Q-team (a group if anonymous people believed to be working with Q), Q research team (a group of Q followers that organize and research the Q-posts) and several other designations (anons, bakers, patriots) that collectively represent the Q-followers. Slogans such as WWG1WGA (Where we go one we go all), and WRWY (We Are With You) are used to reinforce faith and motivate collective action in the QAnon movement. 
    
    \item \textbf{Expectations:} Through the promises of arrests of deep state agents, Q sets expectations for their followers. Expectations are about both good and bad events for the QAnon community. For example, several Q-drops predict arrests of specific deep state politicians while others warn the readers about false flag events, forecasting covert operations of various governments and cabals. 
    
    \item \textbf{Practices:} An important part of QAnon community is hunting for clues provided by Q. Q also instructs their followers to follow certain knowledge construction practices. For example, Q asks their followers to organize and archive Q-posts, connect the dots and dig deeper for the truth. 
    
    \item \textbf{Foes:} Q routinely releases the names of celebrities, politicians and law enforcement agents who are purportedly associated with the deep state. Deep state agents are believed to be involved in a satanic cult with an international child sex trafficking ring. Simply put, foes of QAnon are portrayed as the enemies of the QAnon movement. 
    
    \item \textbf{Heroes:} While the law agencies, media and a large part of the government is considered to be controlled by the deep state, there are some ``good guys'' who fight for the American people according to Q. Q, Donald Trump and former U.S Attorney General Jeff Sessions are at the top of this list.  Heroes often know more than they choose to reveal to the QAnon community for the reasons of national security and are believed to be experts at undercover work.  
\end{enumerate}

As a byproduct of the content analysis process, we recorded relevant words and phrases in each of the five dimensions of the QAnon social imaginaries. We refer to this as a \textbf{seed lexicon} for QAnon social imaginaries. In the next section, we understand how QAnon followers communicate the social imaginaries established by Q, by quantitatively expanding the seed lexicon.

\section{RQ2: Communication of Social Imaginaries by QAnon Followers}
In RQ1, we uncovered five dimensions of the QAnon social imaginaries---\textit{movement}, \textit{expectations}, \textit{practices}, \textit{heroes}, and \textit{foes} put forward by the leader Q. How do QAnon followers communicate these social imaginaries within the QAnon discussion communities? As a result of the qualitative analysis in RQ1, we obtained a seed lexicon of 75 phrases across five dimensions of social imaginaries. For example, see the words in bold in Table \ref{tab:content_analysis_examples} and the words mentioned in Figure \ref{fig:rq1meth} (b). Note that the phrases in the seed lexicon are directly extracted from the Q-drops. However, while discussing the Q-drops, the QAnon followers might adapt various expressions of the phrases. For example, the Q-drops frequently mention ``deep state'' to refer to the collective of allegedly corrupt politicians and celebrities involved in child trafficking rings. However, a manual inspection of the comments in QAnon discussion subreddits revealed that the QAnon followers use various phrases---``swamp'', ``antiq'' (for anti-Q), ``evil pedos''---to refer to the deep state. In this section, we explain our methods for identifying various expressions of phrases in the seed lexicon. As a result, we create the QAnon Canon, a lexicon of 403 phrases capturing the communication practices used by QAnon followers.

\subsection{Method: Creating QAnon Canon}
\begin{figure*}[t]
    \centering
    \includegraphics[width=0.99\textwidth]{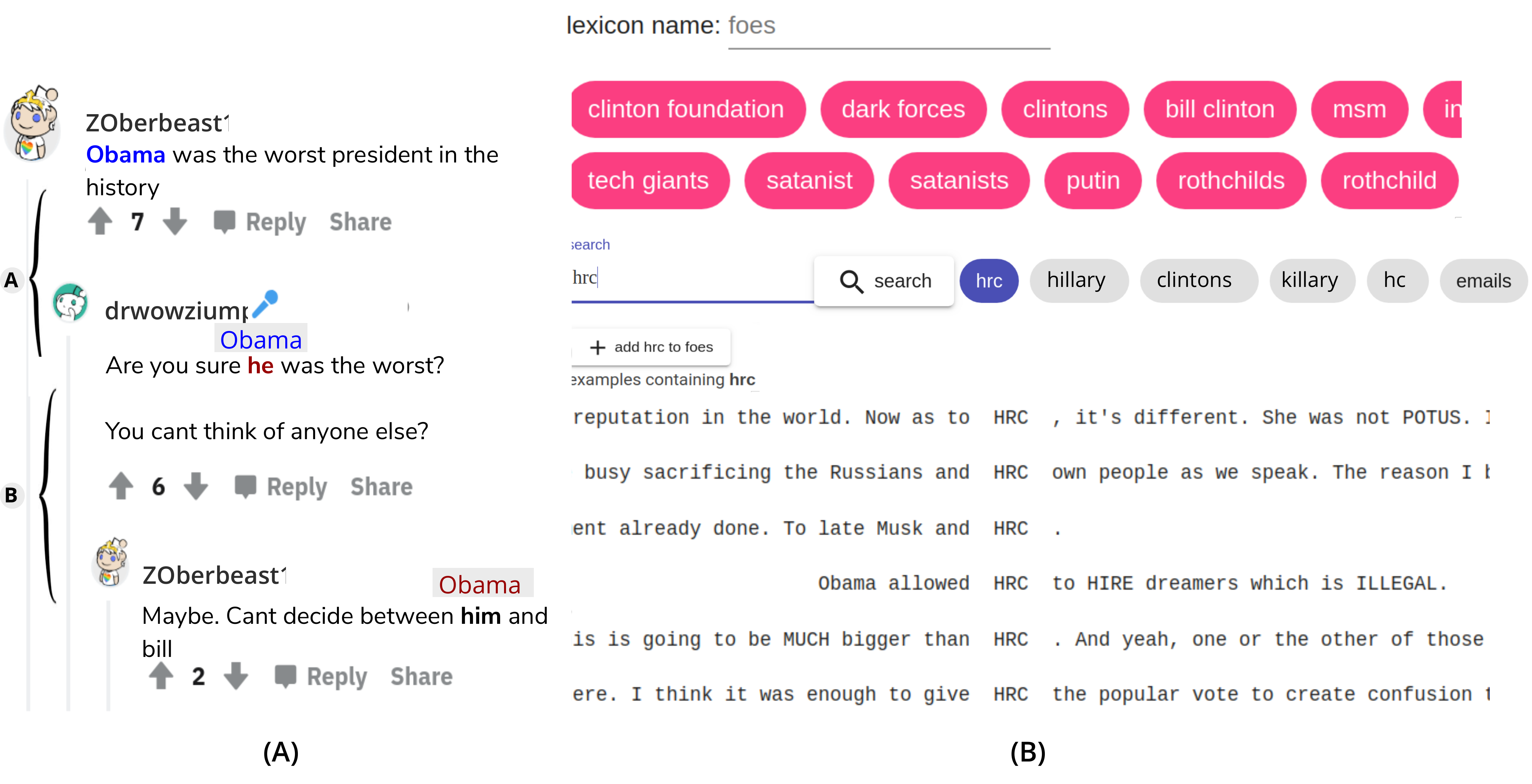}
    \caption{(A) An example discussion thread with coreferences where the entity ``Obama'' from the parent comment is referred by ``he'' and ``him'' in the child comment. To resolve coreferences, we consider the pairs of parent-child comments at a time. This way, with the context of the parent comment, ``he'' in the reply comment will be replaced with the word ``Obama''. (B) A screenshot of the user interface used to iteratively expand the seed lexicon by finding similar phrases. The interface displays pre-loaded phrases in the seed lexicon (pink tabs) which can be searched in the search bar. The search returns top similar phrases (displayed right of the search bar) and the examples of comments containing the searched term. The search bar also has auto-complete feature that suggests lexically similar words to the search term using fuzzy search. Users can add new phrases to the lexicon by either clicking and adding phrases from the search results or manually typing and adding phrases from the examples. The example shows the results for ''HRC''. We select relevant search results (example, hillary, killary, hc) and add them back to the seed lexicon.}
    \label{fig:lex_creator}
\end{figure*}

We quantitatively expand the seed lexicon into QAnon Canon---a dictionary capturing words and phrases in each of the dimension. We use the QAnon Subreddits Dataset of 1.2 M comments and posts to understand various ways in which phrases from the seed lexicon are expressed. We utilize rigorous quantitative methods that use sentence parsing and semantic similarity of words to expand a seed set of 75 phrases to over 403 phrases. Finding various expressions for phrases from public discourse is a challenging task.  Specifically, pronouns are often used to refer to the noun phrases (ex. using `she' instead of `Hillary'). This is called \textit{coreference}. In order to find similar phrases, we first need to resolve the coreferences. After resolving the coreferences, we characterize meanings of various phrases as vectors and use interactive mixed-methods approach to find various expressions of phrases in the seed lexicon. We start by introducing our method for coreference resolution.



\subsubsection{\textbf{Coreference Resolution}} Consider the following Reddit comment: ``HRC is corrupt. She is Evil.'' Here, the first sentence mentions HRC (Hillary Rodham Clinton) and the second sentence refers to the same antecedent entity ``HRC'' by the pronoun ``she''. To derive correct interpretation of this text, we first need to resolve the coreference (HRC---she) where pronouns and other referring expressions must be connected to the right entities. A successful coreference resolution will result in the replacement of the pronoun with correct entity: ``HRC is corrupt. HRC is Evil.'' Coreference resolution is a well explored problem in computational linguistics for studying discourse \cite{stylianou2021neural}.  We use {\small \tt neuralcoref}\footnote{\url{https://github.com/huggingface/neuralcoref}}, with {\small \tt Spacy}\footnote{\url{https://spacy.io/}} pipeline that resolves coreference clusters using neural networks. To further improve the interpretation of a comment, we consider its parent comment while resolving the coreference. For example, see the comment thread in Figure \ref{fig:lex_creator} (a). The first comment mentions an entity ``Obama'' which is referenced with ``he'' and ``him'' in comments in the reply. Hence, while resolving the coreference in a comment, we also need the context of its parent comment. We use breadth first search to first, find pairs of parent--child comments from the top to  the bottom of the comment tree, and then use {\small \tt neuralcoref} to resolve coreferences in the  parent--child comment pair.



\begin{figure*}[t]
    \centering
    \includegraphics[width=0.99\textwidth]{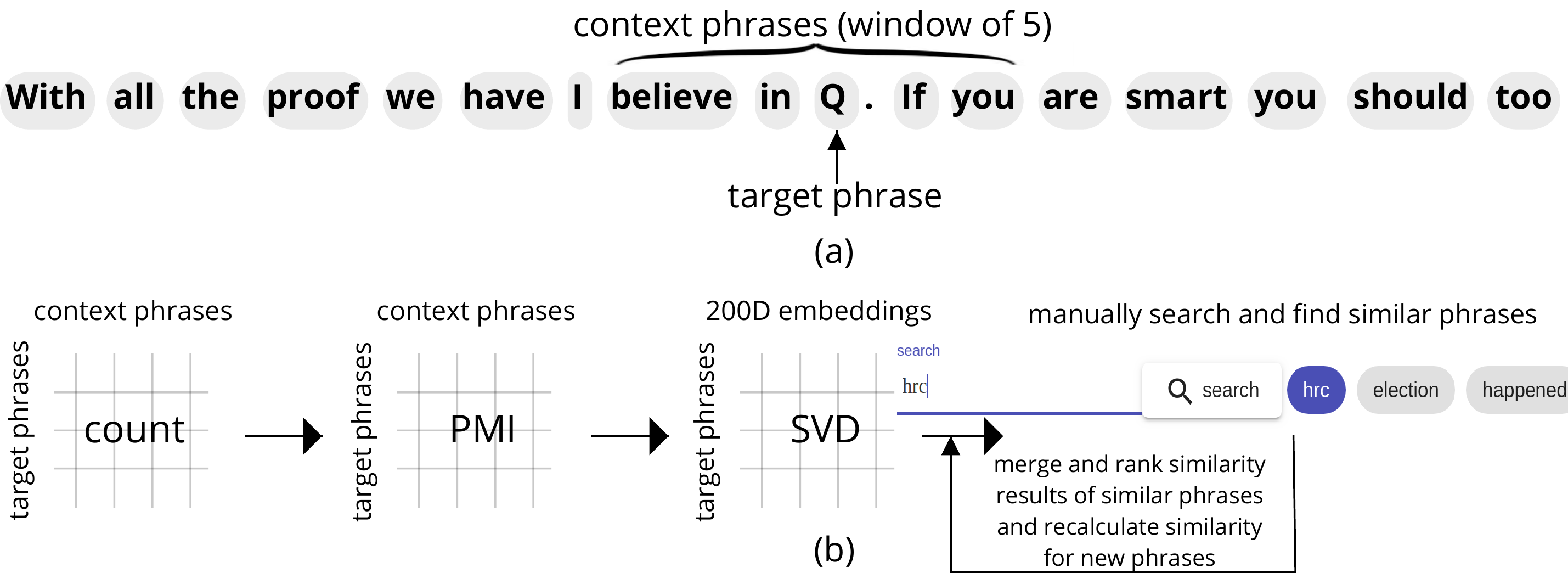}
    \caption{Figure showing the process of characterizing phrase embeddings and updating the embeddings based on similar phrases. (a) For every target phrase, we first record the context phrases based on the co-occurrence in a sentence. (b) Based on the recorded contexts, we build a count matrix of target phrases vs. context phrases. We calculate the pointwise mutual information (PMI) to first encode the phrase similarity and then derive 200 dimensional embeddings for every phrase based on singular value decomposition (SVD). We use the phrase embedding similarity to search for similar phrases in the interface (Figure \ref{fig:lex_creator} (B)) and repeat the process by merging and re-ranking the similarity results of the similar phrases}
    \label{fig:lex_pipe}
\end{figure*}

\subsubsection{\textbf{Characterizing Phrases}} After resolving coreferences, we characterize similarity between different phrases. To find similarity computationally, we first need to represent the meaning of different phrases in the numeric form. This is the process of modeling the meanings of phrases by \textit{embedding} the phrases in the vector space. We first identify phrases by tokening sentences into top unigrams, and bigrams and trigrams using collocations---sequence of commonly occurring words. Note that using collocations allows us to avoid overlap between unigrams and multi-grams. For example, if ``Donald Trump'' is mentioned in a comment, it will get tokenized only as a bigram and not as individual unigrams, ``Donald'' and ``Trump''.  
We create sparse vector representations, or embeddings, for phrases using the information about their co-occurrence with other phrases in the comment. We first calculate phrase co-occurrence matrix which records how often a phrase co-occurs with another phrase. We consider two phrases to ``co-occur'' when they are present together in a comment within a span of five phrases (Figure \ref{fig:lex_pipe} (a)). In other words, to calculate the co-occurring phrases for any particular phrase, we consider the \textit{context} of five phrases. However, raw co-occurrence frequency has several limitations as a measure of association between two phrases. For example, common words like stop words (e.g., ``the'') naturally co-occur with many other phrases according to raw frequency, but provide little information about the words next to them. Hence, we calculate pointwise mutual information (PMI) for every pair of phrases. High PMI between two phrases indicates that the two phrases share a surprisingly higher number of common context phrases \cite{church1990word}. PMI is specifically helpful in surfacing associations to low frequency phrases. However, because the PMI matrix represents associations between every pair of phrases it is high dimensional---a square matrix with number of rows and columns equal to the total number of phrases in the entire corpus. Hence, to obtain computationally affordable and dense embeddings for every phrase, we perform singular value decomposition (SVD) on the PMI matrix (Figure \ref{fig:lex_pipe} (b)). Finally, we capture the semantic meaning of every phrase in a 200 dimensional SVD vector. Now, similarity between the phrases can be obtained by calculating the cosine similarity between their vectors. In the next subsection, we explain our methods for dynamically updating the phrase embeddings to iteratively find similar phrases.

\subsubsection{\textbf{Finding similar phrases}} Consider that we have embeddings for ``Hillary'' and ''HRC''. Given the high similarity between the embeddings, and the manual verification, we conclude that ``Hillary'' and ''HRC'' refer to the same entity. How can we use this information to further find phrases that are similar to both ``Hillary'' and ''HRC''? We use a mixed manual and quantitative approach to iteratively find similar phrases. To ease the process of finding similar phrases, we built a web interface displayed in Figure \ref{fig:lex_creator} (B). 

The web interface shows the seed phrases that belong to the lexicon. It also enables querying for new phrases via a text box that suggests phrase completions as well as similar spellings, and example comments containing the query. Moreover, it shows the 10 phrases that are most semantically similar to the query. In the absence of a query, the interface shows the 10 phrases most similar to the current lexicon. We expanded the lexicon by inspecting and querying for similar phrases that represent related entities and concepts. For example, the seed lexicon for \textit{foes} included ``HRC,'' and ``Hillary'' was automatically suggested as a similar phrase: we added it to the lexicon, which automatically updated the list of similar phrases, and iterated the procedure until no similar phrase belonged to the lexicon. In the process, we noted down related terms that appeared in the example comments, and queried for them, eventually adding them and their similar phrases to the lexicon when appropriate.  

Technically, we compute semantic similarity in two ways. In the case of a query phrase, we compute semantic similarity of a new phrase simply as cosine similarity of the corresponding embeddings. Note that instead of using pre-trained embeddings off the shelf, we use our own trained phrase embeddings to better fit our dataset and similarity context. Using locally generated embeddings also enables us to iteratively modify the embeddings in the interactive lexicon generation phase.
When no query is selected, we compute the 10 most similar phrases to each phrase in the lexicon, merge them into a single list, rank the list according to the similarity to the closest lexicon phrase, and take the 10 highest-ranking phrases.   

Following this procedure, we expand
the lexicon from 75 phrases to 403. We named the final lexicon the QAnon Canon, which we introduce next.



\label{sec:qcanon_results} 
\subsection{Results: QAnon Canon}
\begin{table*}[t]
\centering
\resizebox{0.8\textwidth}{!}{%
\begin{tabular}{@{}ccccc@{}}
\toprule
\textbf{movement} & \textbf{practices} & \textbf{expectations} & \textbf{heroes} & \textbf{foes} \\ \midrule
patriots   & hive mind          & mockingbird      & q                    & kabal       \\
q analyst  & dig the truth      & big drop         & potus                & deep state  \\
q research & treasure hunt      & boom             & white hats           & hillary     \\
anons      & crumbs             & moab             & sessions             & obama       \\
qteam      & clues              & arrests          & wray                 & satanists   \\
wizards    & watch water        & pyramid collapse & mueller              & rothschilds \\
white hats & spider web         & layoffs          & huber                & big pharma  \\
wwg1wga    & future proves past & big news week    & kansas               & red cross   \\
qanon      & trust the plan     & maga promise     & wizards and warlocks & nwo         \\ \bottomrule
\end{tabular}%
}
\caption{Table showing example phrases from the QAnon Canon. In total, the lexicon contains over 403 phrases recorded across 5 dimensions of QAnon social imaginaries. }
\label{tab:canon_example}
\end{table*}

As a result of iteratively finding phrases similar to the seed lexicon, we obtained QAnon Canon---a lexicon of 403 phrases encoding how QAnon followers communicate QAnon social imaginaries. Table \ref{tab:canon_example} displays ten example phrases in each dimension. We are able to recover various expressions for named entities. For example, QAnon Canon contains multiple expressions for Hillary Clinton---hillary, HRC, HC, killary, billary, alice in wonderland, clintons---and Barack Obama---Obama, Hussein, Obamas, ObamaHillaryCIA, Barack, Obummer, 0bama. Similarly we were also able to find lexically and semantically similar expressions of various phrases symbolizing the movement (q-team, q-analyst, q-research, q-clearance), practices (q-drops, qdrops, q drops, qposts, q-posts) and expectations (layoffs, mass exodus). 

We plan to make this lexicon publicly available upon the acceptance of this paper \footnote{\url{https://www.dropbox.com/sh/c284b5xr5r4qrmf/AAClDzwH8f0hvA1D0ZQoWG43a?dl=0}}. In the next research question, we use words from QAnon Canon as linguistic features for identifying expressions of belief and dissonance in the QAnon community.

\section{RQ3: Identifying Expressions of Belief and Dissonance in QAnon}
How do conspiracists express views that are dissonant with the social imaginaries in their communities? To answer, we classify expressions of belief and dissonance in QAnon subreddits. We use various sampling strategies to create a labeled dataset of comments annotated into one of the three categories: \textit{belief} in QAnon, \textit{dissonance} with QAnon and \textit{neutral}. We use this labeled dataset to build an ensemble of machine learning classifiers to identify belief and dissonance in QAnon. We start by introducing our feature set. 


\subsection{RQ3a Method: Compiling Factors in Self-Disclosure of Belief and Dissonance}
Referring to the QAnon Canon and prior literature on expressions of belief and dissonance we compile lexical, stylistic, and document level features for classifying belief and dissonance in QAnon. Our feature set is designed based on related work on expressions of belief and dissonance. We prefer hand-picked, theoretically motivated features over pre-trained sentence embedding models to preserve the interpretability of importance of various features in identifying belief and dissonance. We discuss the importance of different features in Section \mbox{\ref{sec:featimport}}.


\begin{enumerate}
    \item \textbf{QAnon Canon (403 features): } The QAnon Canon lexicon, created in RQ2, captures the social imaginaries of the QAnon community, which can be elicited in affirming belief \cite{arechiga2019mythic}. 
    Similarly, disagreement with the social imaginaries can signal dissonance with the QAnon worldview. Hence, we calculate the frequency of each of the 403 phrases in the QAnon Canon in user comments. 
    \item \textbf{Linguistic Inquiry and Word Count (LIWC) (19 features): } 
    LIWC encodes words that capture affective, emotional and cognitive processing expressions and is often used for analysis of online texts \cite{tausczik2010psychological}. For example, LIWC categories of tentativeness (includes words such as \textit{maybe}, \textit{perhaps}) and certainty (\textit{always}, \textit{never}) were specifically found to be relevant in the expressions of doubts in online reviews \cite{evans2021expressions}. Following a similar rationale, we include 17 other relevant LIWC categories\footnote{feel, discrepancy, second person pronouns, differentiation, religion, third person singular pronouns, causation, first person pronouns, anger, hear, third person plural pronouns, insight, sadness, see, negations, conjunctions, anxiety}. For example, we include conjunctions (and, but, whereas) that may be present in arguments that combine contradictory claims (``I am trying to trust Q \textbf{but} my patience is running out'') \cite{decter2016impressive}. 
    \item \textbf{Integrative Complexity (IC) Score (1 feature): } IC is a psychometric that captures the ability of people to recognize multiple perspectives and connect them together \cite{suedfeld199227}. IC is closely related to expressing belief and attitudes \cite{czaja2016integrative}. IC scores range from 1 to 7 where 1 indicates no evidence of IC and 7 indicates the presence of overarching perspectives with detailed connections. We calculate the IC score using the model published by \citeauthor{robertson2019language} \cite{robertson2019language}.
    \item \textbf{Credibility Cues (8 features): } Perceived credibility and the evaluation of the common knowledge are strongly associated with belief and disbelief \cite{mitra2017parsimonious,pilditch2020false}. We include credibility cues such as booster words (for example, \textit{actually}, \textit{evidently}), hedge words (\textit{in my view}, \textit{in general}), modal words (\textit{hypothetical}, \textit{improbable}) and evidentials (\textit{know}, \textit{guess}) that are associated with perceptions of credibility \cite{mitra2017parsimonious}. Additionally, we also calculate sentiment scores \cite{bordia2004problem} and number of quotations \cite{de2012did} in a comment that might indicate uncertainty, as well as number of questions that might signal information needs \cite{morris1938foundations}. 
    \item \textbf{Community Feedback (2 features): } On Reddit, comments that are well-received by the conspiracy community are awarded upvotes while ill-received comments get downvotes \cite{oswald2021climate}. In particular, we expect comments expressing dissonant views to receive negative community feedback. Hence, we calculate the comment score (an aggregation of upvotes and downvotes). To contextualize the feedback received, we also compute the synchronicity of the comment score with the score of its parent comment. 
    To calculate synchronicity, we subtract the parent comment score from the child comment score.  
    \item \textbf{Generic Document Level features (50 features): } Finally, we also calculate \textit{smooth inverse frequency} (SIF) document embeddings that capture the overall semantics of a comment by combining the embeddings of its words \cite{arora2016simple}. We calculate 50 dimensional SIF embeddings and use them as the baseline for evaluating the features discussed above. 
\end{enumerate}

In total, we calculate 483 features for every Reddit comment or post in the 12 QAnon subreddits. Next, we describe various sampling methods used to create a labeled dataset for classification. 

\begin{figure*}[t]
    \centering
    \includegraphics[width=0.99\textwidth]{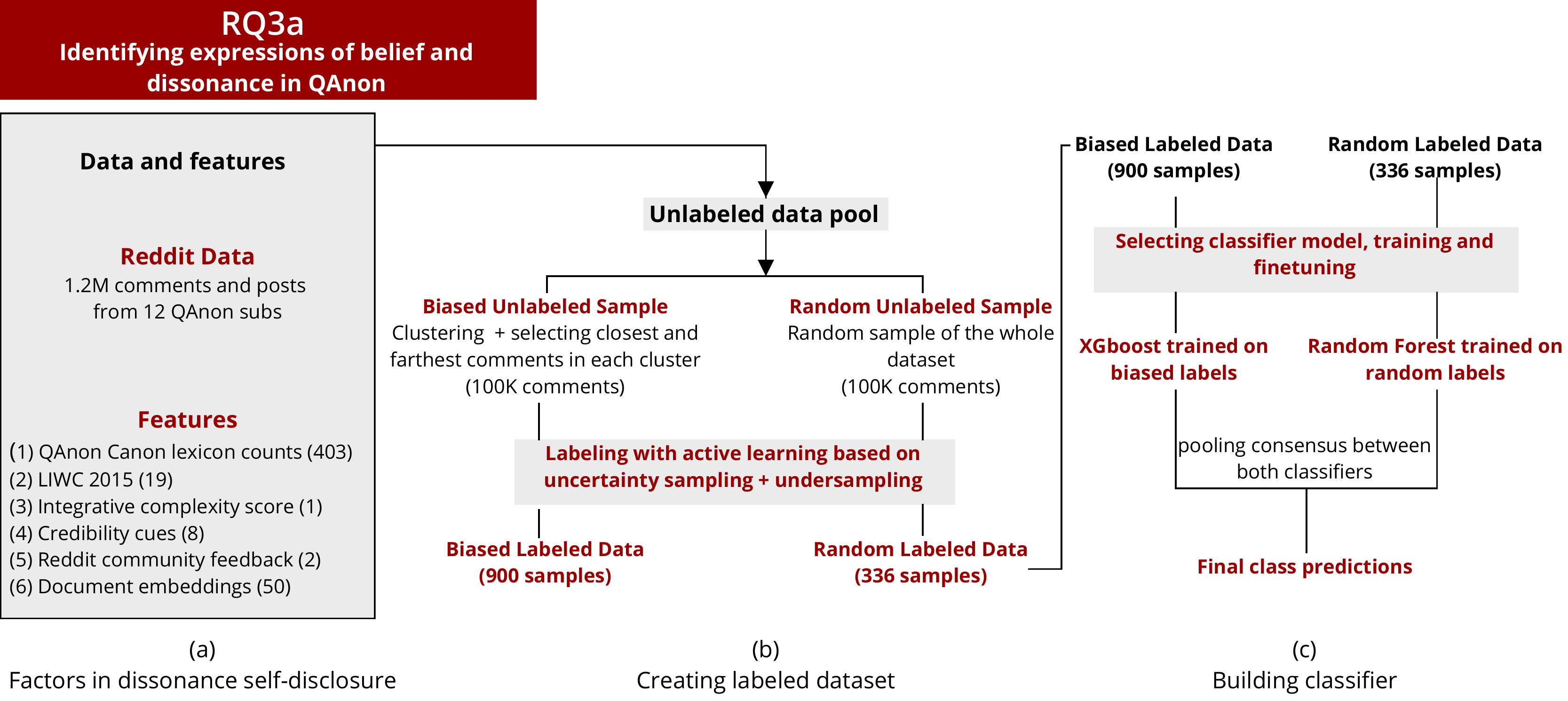}
    \caption{Figure showing RQ3a method flow. (a) First we design and extract features for each of the 1.2M Reddit comments and posts. (b) Due to inherently imbalanced dataset (QAnon subreddits are likely to have very few expressions of dissonance compared to belief or neutral) we use various sampling techniques to potentially find comments expressing dissonance. (c) We use labeled data generated from different sampling techniques to train two machine learning classifiers and consider their consensus as final class prediction.}
    \label{fig:RQ2_method}
\end{figure*}

\subsection{RQ3a Method: Creating a Labeled Dataset}
To understand belief and dissonance at scale, we need a large labeled dataset---ideally, the whole 1.2M comments in the study. We rely on a smaller, high quality labeled dataset that is manually vetted and use a high-precision classifier to extend the labeling to the rest of the data. Yet, even annotating the smaller dataset is challenging: labeling expressions of belief and dissonance in a community like QAnon is a complex and nuanced task which requires sizable theoretical background and expertise with the QAnon social imaginaries. Thus, we cannot rely on crowdworkers for the task. Moreover, self-disclosures of belief and dissonance are rare occurrences, with the great majority of the QAnon subreddits revolving on discussing details of the theories and phatic talk, posing technical challenges to sampling informative instances of belief and dissonance to label. We address these challenges using a expert-in-the-loop mixed-methods approach \cite{dinan2019build}. 

A common strategy to make the most out of constrained labeling resources is active learning. It lets an \textit{interim} classifier choose which next comment would be the most informative if labeled, given the ones already labeled. The intuition behind it is that many comments are similar to each other (e.g., phatic comments making up the majority of the discussions), and labeling multiple instances would not offer a downstream classifier any new information. Instead, labeling a diverse set of comments would better serve the classifier to explore the variety in the whole subreddit. Especially, the comments about which the classifier is the most uncertain at any point, are the ones that, if labeled, would most likely help it discern between classes in the future. Hence, given a pool of comments, active learning selects the most helpful one for an expert to annotate, and re-trains itself adding the newly annotated comment. This procedure repeats until the classifier performance converges, or until the annotation resources (the experts) are exhausted.

Through a pilot annotation, we found that the classes of interest---belief, dissonance, and neutral comments---are extremely imbalanced. Comments expressing disbelief, especially, amounted to only 6\% of the comments in a random sample, while neutral comments amounted to 80\%. Hence, we differentiate the pools of candidate comments to feed into the active learning loop, to trade off between exploring the large variety of comments in the whole dataset, and labeling a meaningful number of belief and dissonance comments. Hence, we select two distinct pools of comments: the first sampled at \textit{random}, to represent data variety; the second sampled with a \textit{biased} strategy to surface a higher number of instances of belief and dissonance. We perform active learning on each pool separately, and then use the comments labeled in both active learning runs to train the final classifier and extend the labeling to the whole dataset. 

For the final classifier, we experimented with different aggregation techniques, from combining the labeled data and training a single classifier, to training classifiers separately on each labeled datasets and combining their predictions into a single score. The latter approach performs best on a held-out validation set. Figure \ref{fig:RQ2_method} outlines the complete labeling and classification pipeline. 
Next, we discuss the details of sampling the random and biased data pools, performing active learning, and training the final classifier.

\subsubsection{\textbf{Creating Pools of Unlabeled Data}}
Because of the inherent disproportion of neutral comments compared to those expressing belief and dissonance in the QAnon subreddits, classifiers trained on such data may be less accurate on the latter classes \cite{attenberg2010label} even when trained on large labeled data \cite{c2018active}. 
In order to accurately model belief and dissonance in QAnon, we need a higher proportion of labeled instances of such classes. At the same time, to build a classifier generalizes well to real data distributions, we also need labeled samples that represent the overall dataset. Hence, we first create two unlabeled sample pools---\textit{random samples} and \textit{biased samples}.

\begin{enumerate}
    \item \textbf{Random Unlabeled Sample:} Random sample contains 100K comments and posts selected uniformly at random. Random sampling can be representative of the dataset \cite{lance2016sampling} where various types of expressions of belief and dissonance can occur at their natural frequencies inside the QAnon subreddits. 
    \item \textbf{Biased Unlabeled Sample:} We use a cluster-based sampling technique to include belief, dissonance and neutral expressions in similar quantities \cite{yuan2011initial}. Specifically, we perform K-Means clustering on the entire dataset and select samples that are closest and farthest from each cluster centroid. We use the elbow method---plotting explained variance in clustering as a function of number of clusters--to determine optimal number of clusters as 3. From every cluster, we collect the 20K closest and farthest samples to each centroid, for a total of 120K posts and comments.
\end{enumerate}

We use these two pools of unlabeled data to select the samples to label. 

\subsubsection{\textbf{Labeling Dataset with Active Learning: }}

\begin{figure*}[t]
    \centering
    \includegraphics[width=0.99\textwidth]{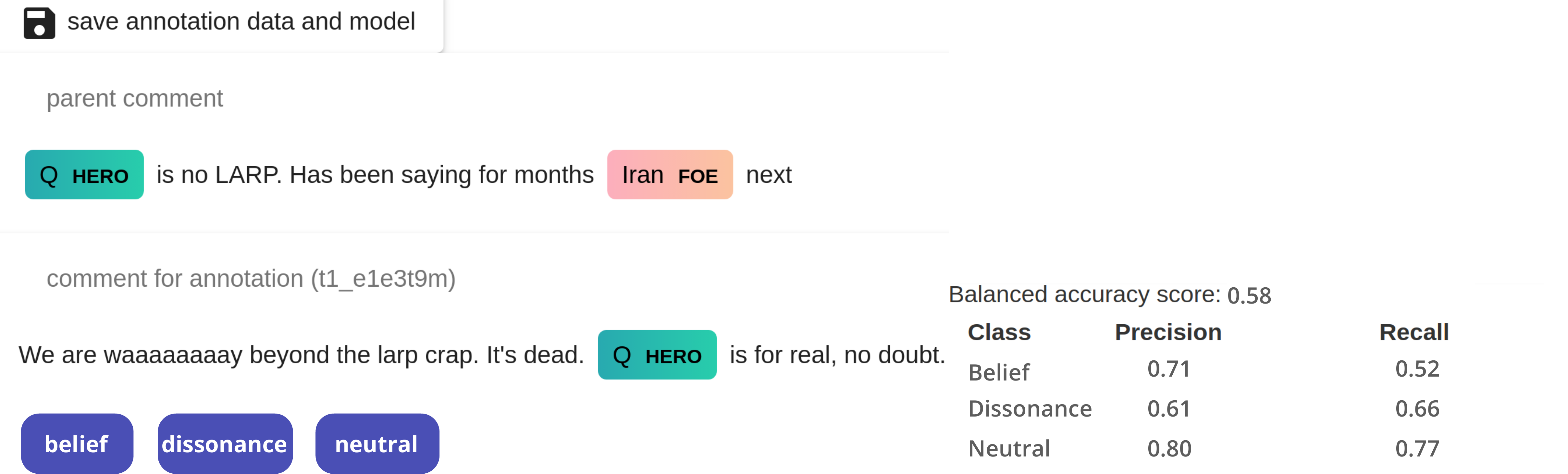}
    \caption{Figure showing RQ2 annotation interface. Every comment and its parent comment is annotated with categories of QAnon social imaginaries. Annotators label every comment as belief, dissonance or neutral. The interface also displays naive classification scores updated every 10 samples.}
    \label{fig:RQ2_interface}
\end{figure*}
Combined, the random and biased data contains 220K comments, which are still too large for complete labeling. To trade-off between the manual labor of labeling many comments and the downstream classification performance which depends on a large enough labeled dataset, we devise an interactive labeling process based on active learning. An active learning classifier selects the next unlabeled comment to label according to its sampling strategy; after annotators label the comment, the classifier adds the comment to the labeled dataset, retrains itself, and selects the next comment to label \cite{olsson2009literature}. 
This process repeats until the satisfactory classification performance is achieved. The crux of the learning strategy lies in sampling the new data to be labeled. We use a popular sampling strategy: uncertainty sampling.

\textbf{Uncertainty Sampling for Active Learning: }
A common strategy for finding the best instances to label is to choose those unlabeled instances the classifier is currently most uncertain about. The intuition is that such instances would add the most information to the labeled dataset, and would therefore tighten the classification margin in the fewest iterations. One measure of uncertainty is entropy---a general measure of disorder in a system. In the context of classification, high entropy of predicted class probabilities indicates higher uncertainty of the classifier \cite{holub2008entropy}. 
Based on the probability distribution $P (y_{i}|\mathcal{X})$, where $y_{i}$ is the predicted probability of class $i$ and $\mathcal{X}$ is the sample, entropy for each sample is simply calculated as:

\begin{equation} 
    entropy(\mathcal{X}) = - \sum_{i=1}^{C} P (y_{i}|\mathcal{X}) log_{2}(P(y_{i}|\mathcal{X}))
\end{equation}

\textbf{Interactive Labeling with Active Learning: } Figure \ref{fig:RQ2_interface} showcases the web interface we built to facilitate interactive labeling with active learning. The interface displays the comment for annotation along with its parent comment for the context. Within the comments, phrases belonging to the QAnon Canon are highlighted and tagged with the dimension which they belong to. This helps annotators consider QAnon social imaginaries while identifying expressions of \textit{belief} and \textit{dissonance}, and to determine whether a comment is \textit{irrelevant}. 
The first two authors of the paper annotated the whole labeled dataset.
We use labeling guidelines detailed in Appendix \ref{sec:codebook} to label the samples. 
The active learning classifier estimates its performance in cross-validation and on a held-out test set every 50 labeled samples. Accordingly, the interface also displays the current change-adjusted balanced accuracy (random performance scores 0, perfect performance 1) and precision and recall for all three classes. We continue the annotation process until precision and recall surpass 0.60 for all classes. Through this process, we labeled 1,204 comments from the random sample and 1,167 comments from the biased sample. 

\subsection{RQ3a Method: Building Classifiers from Labeled Data}
We use this labeled data to develop a classifier that reliably identifies expressions of belief and dissonance in the whole QAnon subreddits. In this section, we explain our selection of the classifier model, and details about its training and prediction procedures. 

\subsubsection{\textbf{Selecting the Classifier Model: }}
Although we labeled a total of 2,371 comments and posts from the random and biased samples, the labeled dissonance instances were still fewer compared to the belief and irrelevant. Since class imbalance may bias classifier performance, we balance the data. We undersample the data while maintaining above 0.6 precision for all classes. We experimented with various undersampling strategies and found the best performance with one-sided undersampling \cite{kubat1997addressing}. One-sided sampling removes the noisy, under-performing examples from the majority classes while preserving all the examples of the minority class. After undersampling, we end up with  336 and 900 samples from the random and biased labeled datasets respectively.

To choose best classifier we use the {\small \tt auto-sklearn}\footnote{\url{https://github.com/automl/auto-sklearn}} toolkit that searches over wide variety of models optimizing for performance \cite{feurer2020auto}. We find the optimal model for each of the two labeled datasets---namely, a Random Forest model \cite{ho1995random} for the random labeled data and an Extreme Gradient Boosting (XGBoost) model \cite{chen2016xgboost} for the biased labeled data. 

\subsubsection{{\textbf{Training and Predicting with Classifier Models:} }}
We perform hyperparameter optimization to tune model parameters. Different models require tuning of internal parameters such as learning rate, estimators, etc., to generalize to unseen data \cite{claesen2015hyperparameter}. We determine the best model parameters through an extensive grid search in a cross-validation scheme. We test over 15000 combinations of hyperparameters, optimizing for chance-adjusted, balanced classification accuracy. 

After fine-tuning both classifiers, we combine their predictions on the whole dataset. Specifically, we use a strict consensus pooling method to determine class assignments. Consensus pooling assigns a particular class (belief, dissonance or neutral) to a sample if \emph{both} classifiers agree on the predicted class. Disagreements are removed from the predictions. We also experimented with different pooled prediction strategies that do not require removing the ambiguous samples (see Appendix section \ref{sec:pooling_strats}) with slightly lower scores. A single classifier trained on the combination of the two labeled datasets performs significantly worse.

\subsection{RQ3b Method: Characterizing the types of Dissonance Self-Disclosures}
Using the classifiers designed in the previous steps, we label dataset of 1.2M comments and posts. We remove around 500K samples with prediction disagreements. In the remaining automatically labeled dataset, we find that over 43K comments and posts (~6\%) are labeled as dissonant. What are the different ways in which users express dissonance? 
We qualitatively analyze a random sample of 500 comments and posts expressing dissonance, focusing on how dissonance relates to the dimensions of QAnon social imaginaries, such as how they refer to collective practices and expectations. 
We report the results of the qualitative analysis in the section \ref{sec:reesult_RQ3b}.

\begin{table*}[t]
\centering
\resizebox{\textwidth}{!}{%
\begin{tabular}{@{}ccccc|cccc|cc@{}}
\toprule
 &
  \multicolumn{4}{c|}{\textbf{Random Forest (RF) Classifier}} &
  \multicolumn{4}{c|}{\textbf{XGBoost Classifier (XGB)}} &
  \multicolumn{2}{c}{\textbf{RF+XGB Consensus}} \\ \midrule
\textbf{} &
  \multicolumn{2}{c}{training} &
  \multicolumn{2}{c|}{validation} &
  \multicolumn{2}{c}{training} &
  \multicolumn{2}{c|}{validation} &
  \multicolumn{2}{c}{validation} \\
\textbf{} &
  precision &
  recall &
  precision &
  recall &
  precision &
  recall &
  precision &
  recall &
  precision &
  recall \\
\textbf{belief} &
  0.61 &
  0.46 &
  0.55 &
  0.41 &
  0.60 &
  0.54 &
  0.66 &
  0.58 &
  0.71 &
  0.54 \\
\textbf{dissonance} &
  0.62 &
  0.76 &
  0.52 &
  0.69 &
  0.60 &
  0.59 &
  0.60 &
  0.64 &
  0.70 &
  0.76 \\
\textbf{neutral} &
  0.71 &
  0.71 &
  0.67 &
  0.69 &
  0.64 &
  0.61 &
  0.70 &
  0.73 &
  0.79 &
  0.79 \\
\textbf{\begin{tabular}[c]{@{}c@{}}balanced \\ accuracy\end{tabular}} &
  \multicolumn{2}{c}{0.66} &
  \multicolumn{2}{c|}{0.60} &
  \multicolumn{2}{c}{0.63} &
  \multicolumn{2}{c|}{0.69} &
  \multicolumn{2}{c}{0.79} \\ \bottomrule
\end{tabular}%
}
\caption{Training and validation performances for individual classifiers and final consensus classifier. The final consensus classifier clearly outperforms the individual classifiers across all precision scores.}
\label{tab:train_val}
\end{table*}

\subsection{RQ3a Results: Classification Performance}
Table \ref{tab:train_val} shows the training and validation performance of the individual classifiers as well as the ensamble classifier consensus-pooling their predictions. Individual classifiers have comparable performances. RF shows 0.66 accuracy in the training set (cross-validated) and 0.60 accuracy on a held-out validation test, while XGB 0.63 and 0.69 respectively. The consensus classifier clearly outperforms both RF and XGB individually. In particular, precision exceeds 0.7 for all three classes. 
We further include distribution plots for validation precision and recall over 100 training iterations in Figure \ref{fig:validation} in the Appendix. The distributions show the mean precision/recall values along with the standard deviations. Overall our results show relatively stable precision and recall values, lessening potential concerns of model overfitting.

\subsection{RQ3a Results: Important Indicators of Belief and Dissonance}
\label{sec:featimport}

\begin{figure*}[t]
    \centering
    \includegraphics[width=0.7\textwidth]{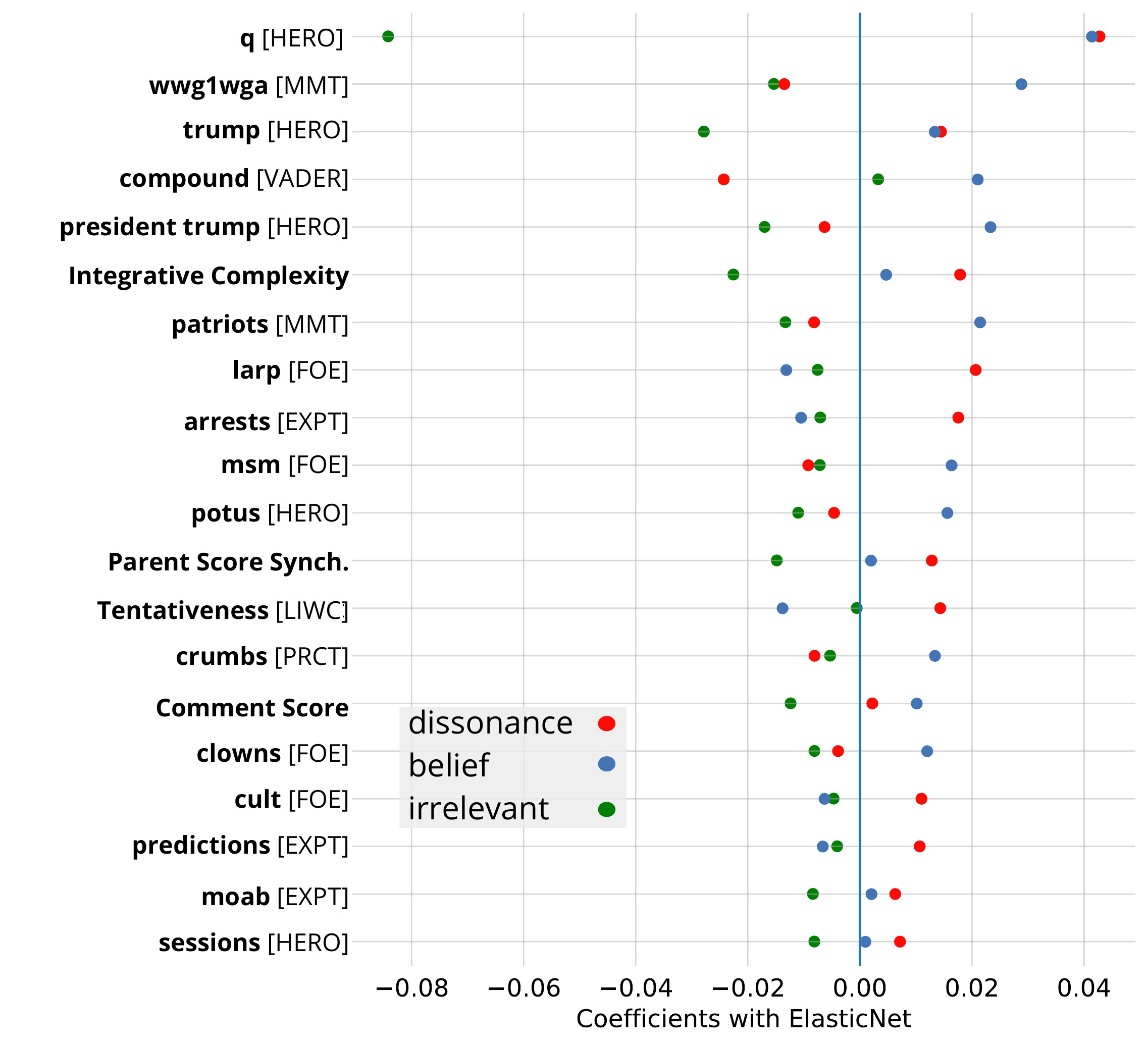}
    \caption{Plot showing most predictive features for each class. Negative coefficients indicate that the low value of feature is associated with the presence of the class. For example ``wwg1wga'', a QAnon movement related word, is present more in the comments expressing belief (0.03 coefficient for belief) whereas dissonance and neutral comments have less occurrences of the same word (-0.015 for both). 
    Note that the importance of features was separately calculated for each class using a  Multitask Elastic Net model. Hence, the coefficients for features for each class are independent of the others. We annotate each feature with the feature category whenever applicable. Abbreviations for QAnon Canon features are; HERO:heroes, FOE: foes, MMT: movement, EXPT: expectations, PRCT: practices. }
    \label{fig:elasticnet_importance}
\end{figure*}

Which features are most impactful in classifying belief or self-disclosures of dissonance? Usually, important features in the model can be interpreted using classification coefficients. Such model explanations are easiest in binary classification where one coefficient can be interpreted in terms of either of the two classes. However, our classification consists of three classes: belief, dissonance and neutral. We need to understand how are different features impactful in classifying \emph{each} of the classes. Hence, to understand the feature importance, we use Multitask Elastic Net model. The multi-task model solves three regression problems with three classes while sharing the same feature space. Hence, we can get feature importance separately for every class. For example, positive feature coefficients for belief class does not imply negative coefficient for the dissonance class. 
Figure \ref{fig:elasticnet_importance} displays most important features for each class. Negative coefficients indicate that the low value of feature is associated with the presence of the class. For example, mention of ``Q'' is positively associated with both, belief (0.04) and dissonance (0.04) whereas, neutral comments have low mentions of ``Q'' (0.08). Interestingly, while both belief and dissonance have higher mentions of ``trump'' (0.01 and 0.01), only comments with belief mention ``president trump'' (0.02) indicating respect for QAnon \textit{heroes}. Comments expressing belief also have higher proportion of \textit{movement} related words (``wwg1wga'' (0.03), ``patriots'' (0.02), ) and higher positivity (compound VADER (0.02)). Dissonance disclosures however, mention more \textit{expectations} related words (``arrests'' (0.02), ``predictions'' (0.01), ``moab'' (0.01)). Dissonance self-disclosures also have higher integrative complexity (IC) score indicating the presence of multiple argumentative perspectives. In general, we find that semiotic language captured by QAnon Canon is strongly indicative of belief (``wwg1wga'', ``president trump'', ``patriots'', ``msm'', ``potus'', ``crumbs'', ``clowns'', ``obummer'' etc.) and dissonance (``larp'', ``arrests'', ``cult'', ``sessions'', ``moab'' etc.). 


\begin{table*}[]
\centering
\resizebox{\textwidth}{!}{%
\begin{tabular}{cl}
\multirow{2}{*}{\textbf{\begin{tabular}[c]{@{}c@{}}Legitimacy of \\ the movement\end{tabular}}} &
  \textit{\begin{tabular}[c]{@{}l@{}}...training your FOLLOWERS to put your voice before Q's - as we see from some - that can definitely be a problem\\ ....this makes me question if this whole thing is even worth it..\end{tabular}} \\ \cline{2-2} 
 &
  \textit{\begin{tabular}[c]{@{}l@{}}i knew this movement was going to shit when the ``Cult of Q'' occupied this sub. \\ YOU NEED TO CHILL THE F' DOWN\end{tabular}} \\ \hline
\multirow{2}{*}{\textbf{\begin{tabular}[c]{@{}c@{}}Unfulfilled \\ expectations\end{tabular}}} &
  \textit{\begin{tabular}[c]{@{}l@{}}Fool me once shame on Q, fool me twice.. Been following Q from beginning. \\ How many failed predictions does it take before you realize its bs?\end{tabular}} \\ \cline{2-2} 
 &
  \textit{\begin{tabular}[c]{@{}l@{}}Ok, I'm getting off the Q-Anon train right here.  He has consistently said, "Big news week", "Past Unlocks the Future", \\ "Have Faith, Patriots are in Control", but nothing ever happens ! We have all been strung along like lemmings.\end{tabular}} \\ \hline
\multirow{2}{*}{\textbf{\begin{tabular}[c]{@{}c@{}}Ineffective\\ practices\end{tabular}}} &
  \textit{this is just ridiculous. Q sends us down the rabbit hole, asks us to ``dig deeper'' but nothing makes sense!} \\ \cline{2-2} 
 &
  \textit{\begin{tabular}[c]{@{}l@{}}In all fairness, Q has only dropped vague crumbs and we are supposed to find the truth off of that? \\ Is it too much to ask for more proof?\end{tabular}} \\ \hline
\multirow{2}{*}{\textbf{\begin{tabular}[c]{@{}c@{}}Distrust in\\ heroes\end{tabular}}} &
  \textit{\begin{tabular}[c]{@{}l@{}}I'm seriously starting to doubt Q is a White House insider. \\ Most of what he posts is easily found in the news and on conspiracy sites.\end{tabular}} \\ \cline{2-2} 
 &
  \textit{\begin{tabular}[c]{@{}l@{}}Trump staff found holding devils signs in their hand. \\ I think trump is controlled too by the jews/nwo..\end{tabular}} \\ \hline
\multirow{2}{*}{\textbf{\begin{tabular}[c]{@{}c@{}}Trust in \\ foes\end{tabular}}} &
  \textit{\begin{tabular}[c]{@{}l@{}}Without more evidence than a few vague posts from Q, \\ I'm not going to believe that Barack Obama was sexually abusing that girl.\end{tabular}} \\ \cline{2-2} 
 &
  \textit{\begin{tabular}[c]{@{}l@{}}I disagree with Q on the AJ {[}Alex Jones{]} matter. I have listened to infowars for years. \\ What's the point in alienating people like Alex who are clearly on our side\end{tabular}}
\end{tabular}%
}
\caption{Types of dissonance related to social imaginaries in QAnon. In RQ3b, we conducted qualitative analysis of predicted dissonant comments. This table presents examples related to main observations. }
\label{tab:rq3b_results}
\end{table*}

\subsection{RQ3b Results: Points of Dissonance in QAnon}
\label{sec:reesult_RQ3b}
In the qualitative analysis of the dissonant comments, we examined how dissonance is expressed along the social imaginaries of QAnon. Table \ref{tab:rq3b_results} lists various points of fracture in the QAnon social imaginaries along with the example comments. Several users express dissatisfaction with various components of the QAnon movement. For example, some comments expressed how YouTubers profiting off of QAnon movement could harm the unity. Moreover, some users expressed concerns with overzealous nature of and the credibility of the others in the QAnon movement. Next, disappointment over Q's failed prophecies and unfulfilled promises is one of the primary point of dissonance for some users. Several users refer to specific phrases used by Q to express dissatisfaction over unmet expectations. While some users doubted the effectiveness of knowledge construction practices instructed by Q, distrusting the legitimacy and power of heroes is perhaps the most common point of dissonance in the QAnon followers. Some users express concerns over the mysterious identity of Q. We did not find many comments explicitly defending the enemies of the QAnon movement however, some users expressed concerns over lack of evidence for vilifying deep state politicians. Some users also expressed disappointment when Q deligitimized popular right wing celebrities such as Alex Jones.

\section{RQ4: User Engagement after Dissonance Self-Disclosure}
Dissonance can induce various behavioral changes, such as strengthening commitment \cite{festinger1959cognitive}, recruiting others \cite{festinger1962theory} or even reversing one's belief \cite{festinger1959cognitive,mcgrath2017dealing}. 
Thus, we study how users change their engagement patterns within the QAnon subreddits after expressing dissonance. 

\subsection{RQ4 Method: Observing Changes in User Engagement After Dissonance}
We use Interrupted Time Series (ITS) analysis to characterize user contributions before and after dissonance self-disclosure. ITS is a quasi-experimental statistical method that is used to analyze change in longitudinal data after an intervention or policy change. For example, researchers have used ITS to observe how dramatic events change user engagement in Reddit conspiracy communities \cite{samory2018conspiracies}. Here, we consider dissonance self-disclosure as the \textit{intervention} that determines changes in user contributions. With ITS, we can characterize whether, and by what degree, the trends in contributions after the intervention differ significantly from before. ITS analysis involves solving the linear regression:

\begin{equation}
    contributions  \sim  b_{0} + b_{1}T + b_{2}D + b_3{P}
\end{equation}

where $T$ represents the time step of the observation, $D$ is a binary variable representing whether the time step is before or after the intervention and $P$ encodes the time steps after the intervention. For example, in a model with time step of a week and an observation window starting 2 weeks prior to the intervention, the 2nd week after the intervention will be encoded as $T=5$ (5 weeks since start of the observation window), $D=1$ (after intervention), $P=3$ (3 weeks after intervention, including the week of intervention). $b_{0}$ is the model intercept. The coefficient of $T$, $b_{1}$ indicates the slope of trend in the outcome variable (contributions in this case) \emph{before} the intervention. $b_{2}$ indicates the change in level \emph{starting at} the intervention whereas $b_{3}$ indicates the change in slope \emph{after} the intervention. Therefore, the actual slope of trend after the intervention is derivable adding $b_{1}$ to $b_{3}$. The ITS regression directly indicates whether the pre-intervention trend $b_{1}$ and change in level at the intervention $b_{2}$ are statistically significant. Since ITS does not model directly the actual slope of trend after the intervention, but only the change with respect to the trend before, we corroborate its statistical significance via a separate piece-wise linear regression.

\subsubsection{\textbf{The ITS Setup: }} 
Given that the QAnon communities were banned within 11 months of their creation, a week is an appropriately short time-step to measure the immediate effects of the intervention. We define an observation window of total 13 weeks, centered at the intervention \footnote{we also experimented with observation windows of 5, 7, 9, 11, 15, 17 and 21 weeks observing similar results.}. We compute the number of contributions (comments and posts) that users made each week in the QAnon communities, normalized by the number of contributions throughout their lifespan. In other words, each observation shows what fraction of contributions users made in QAnon communities within that specific week. 
Normalizing this way allows us to compare all users' contributions from the scale of 0 to 1. We employ several other robustness measures to ensure that users with inherently short contribution spans do not influence the analysis. For example, we consider only users who have at least one contribution before and after the 13-week observation window and also at least one contributions before and after the intervention within the observation window. 
We compare how user engagement changes after expressing dissonance within and outside the QAnon community, by repeating the ITS analysis but including contributions made outside of the QAnon subreddits. As a further point of comparison, we also consider belief, instead of dissonance, as an intervention which may determine changes in engagement within the QAnon subreddits. 

\subsection{RQ4 Method: Correlating Dissonance and Belief with Tenure in the Community}
Next, we turn to the question: how do self-disclosures of belief and dissonance affect users' permanence in the QAnon community? We set out to analyze the self-disclosures that users perform in their first 100 comments, and use them to predict their long-term tenure. The subreddits that host the community were active at different times (Figure \ref{fig:qsubs}); this may confound the analyses because users may stop posting either as a consequence of their experiences of dissonance, or because the subreddit they primarily posted in became banned. Therefore, we restrict the analyses to users who post on the subreddit \texttt{r/greatawakening}, which is the largest and longest running in our dataset. Next, we limit to users who contributed more than 100 comments to the subreddit, and characterize the disclosures of belief and dissonance in their first 100 comments. We compute the total number of disclosures of belief and dissonance as regressors. Moreover, to capture the relationship between belief and dissonance, we compute the dissonance index $\mathcal{D}$ as proposed by \citeauthor{festinger1962theory} \cite{festinger1962theory}. The dissonance index corresponds to the fraction of disclosures that is dissonant:
\begin{equation}
\label{eq:dissonance}
    \mathcal{D} = \frac{dissonance}{dissonance+belief}
\end{equation}
Thus, we add to the list of regressors, the average and maximum dissonance index $\mathcal{D}$. Moreover, as control variables, we compute the minimum, average, and maximum score of their comments, as well as the time of their first and 100th post: users may leave the community because of negative feedback by their peers, or because they post only infrequently to the subreddit. We standardize regressors and control variables, and use them as factors determining future user permanence. We consider the last post in the subreddit as the time when the user leaves the community. To avoid confounds, we limit our observations to users who left the community at least one month before the subreddit were banned (a practice known as censoring in survival analysis). We predict the number of remaining days that the user will spend on the subreddit; to this end, we use negative binomial regression which is suitable for ordinal outcomes (see Table \ref{tab:nbdeath} in Appendix). We also predict the number of remaining comments using ordinary least squares regression (log-scaled to account for skew, see Table \ref{tab:olsdeath} in Appendix), and the binary outcome of whether the user will remain in the community for more than 10 days via logistic regression (see Table \ref{tab:logitdeath} in Appendix). We report results for the subset of regressors that produce the best model fit.

\subsection{RQ4 Results: Changes in User Contributions after Dissonance} \label{sec:rq4results}
\begin{figure*}[t]
    \centering
    \includegraphics[width=0.8\textwidth]{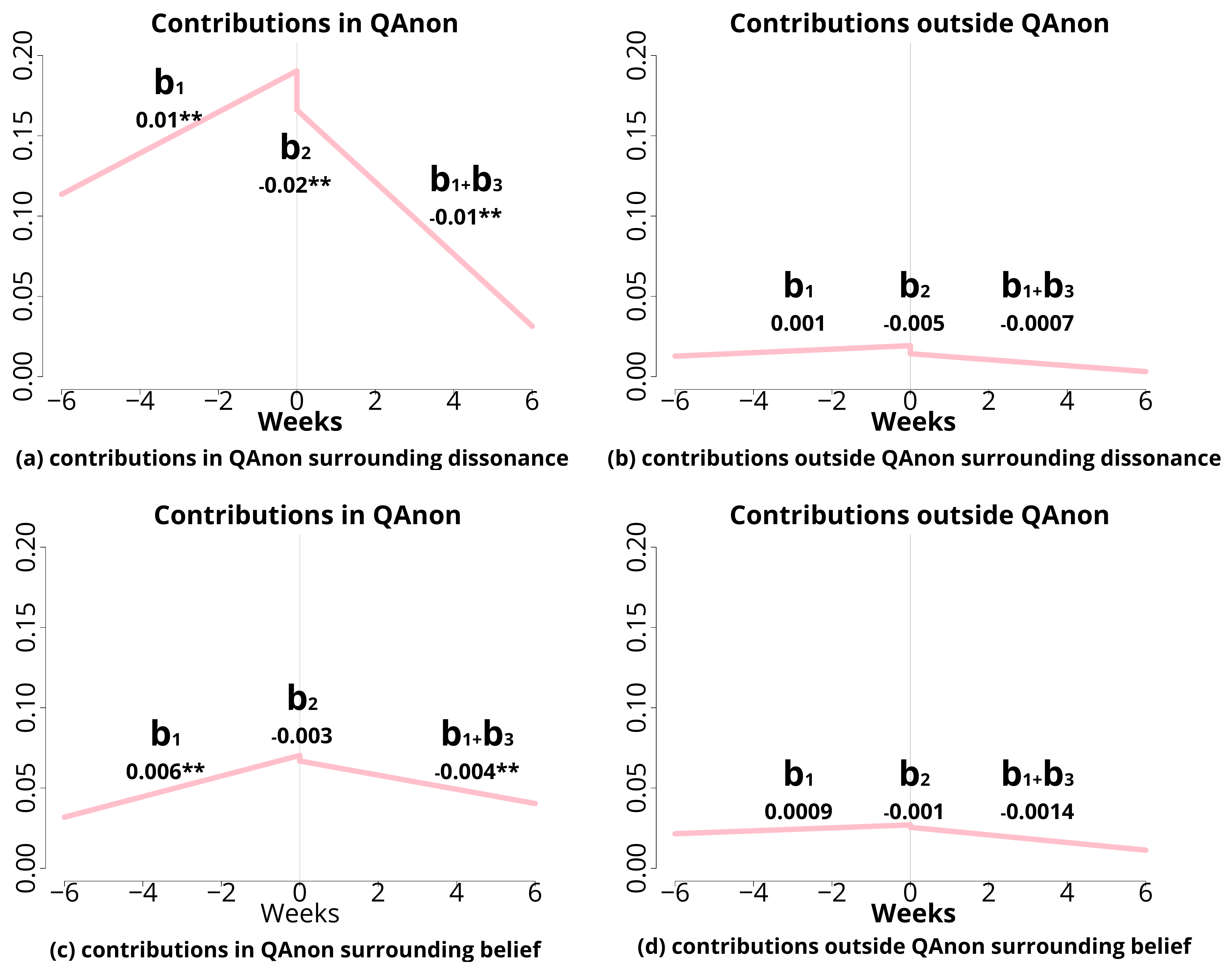}
    \caption{ITS plots for user contributions with different interventions. Two asterisks (**) indicate that the coefficient is statistically significant ($p < 0.05$). As shown in (a), immediately after the dissonance (week=0) there is significant decrease in the user contributions inside QAnon subreddits ($b_{2}$ = -0.02). However, as indicated in (c) there is no significant change in contributions after expressing belief. In the long term, contributions inside QAnon decrease with higher rate after expressing dissonance ($b_{1}$ + $b_{3}$ = -0.01) compared to belief ($b_{1}$ + $b_{3}$ = -0.004). Moreover, (b) and (c) indicate that there are no significant changes in user contributions outside of QAnon after expressing dissonance or belief. }
    \label{fig:its_labeled}
\end{figure*}

We identify users who express dissonance in the manually labeled dataset to perform ITS analysis. We rely on the manually labeled dataset, rather than the larger but automatically labeled one, to be completely confident in our identification of dissonant comments and by association, dissonant users. Figure \ref{fig:its_labeled} displays the ITS and regression results. We find that user contributions inside QAnon decrease significantly ($b_{2}$ = -0.02) immediately after expressing dissonance. The same effect is \textit{not} significant after the the users express belief. Further, in the long term as well, contributions inside QAnon decrease at a higher rate after expressing dissonance ($b_{3}$ = -0.02) compared to expressing belief ($b_{3}$ = -0.004). Is this effect a byproduct of users reducing their overall Reddit activity? 
This appears not to be the case: we find no significant changes in the user activity outside the QAnon subreddits (Figure \ref{fig:its_labeled} (b) and (d)). While the analysis in Figure \mbox{\ref{fig:its_labeled}} is based on  the 2,371 manually labeled comments spanning over 1,498 users, we also repeat the entire ITS analysis on the complete dataset of 700K comments with labels predicted by the classifier. We find similar results indicating that user contributions decrease significantly soon after dissonance (Figure \ref{fig:its_predicted} in Appendix).

\subsection{RQ4 Results: Dissonance and Belief Predict Departure from the Community}\label{sec:rq4departure}

Since users reduce their participation after expressing dissonance, we test whether self-disclosures of dissonance and belief ultimately lead to the users leaving the community. We present here the three models predicting the number of remaining days that the user will spend on the subreddit (Table \ref{tab:nbdeath} in Appendix), the number of comments that the users will contribute in the future (Table \ref{tab:olsdeath} in Appendix), and whether the user will remain in the community for more than 10 days (Table \ref{tab:logitdeath} in Appendix). All three models indicate that a higher number of self-disclosures of dissonance correlates with a shorter tenure. There appears to be an asymmetric effect: whereas dissonance correlates with users leaving, belief does not correlate with them staying longer. While disclosures of belief are not significant correlates per se, it is important to consider them in combination with disclosures of dissonance: the maximum dissonance index $\mathcal{D}$ (Eq.~\eqref{eq:dissonance}) experienced in the first 100 comments significantly correlates with users leaving the community in the following 10 days. These results support and extend those in section \ref{sec:rq4results}, showing that not only disclosures of disbelief are followed by a decrease in contributions, but also by the departure of the users from the community.

\section{Discussion}
In this paper we uncover the dimensions and expressions pertaining to social imaginaries in a conspiracy community. Utilizing the expressions of social imaginaries, we identify how conspiracy community members express their belief and dissonance towards conspiratorial views discussed in the community. Further investigation of their dissonant communication, allows us to outline the points of fracture in their conspiracy belief system. 

In RQ1 and RQ2, our analysis yielded novel dimensions and language correlates along which the QAnon conspiracy community aligns their social imaginaries. This led to characterizing QAnon social imaginaries along the conceptual dimensions of \textit{movement}, \textit{practices}, \textit{expectations}, \textit{heroes}, \textit{foes}, and to formalizing the symbolic language used in conveying shared meanings across each dimension. We find that various such symbolic words are indeed important indicators of belief and dissonance in the QAnon community. For example, the phrases symbolizing QAnon \textit{movement} such as ``WWG1WGA'' (where we go 1 we go all) and ``patriots'' are the top predictive features of belief expressions. 
Similarly, words related to \textit{expectations} such as ``arrests'', ``predictions'', and ``moab'' are indicative of dissonant expressions.  
Moreover, we find that self-disclosures of belief and dissonance are consequential to understanding engagement in the QAnon conspiracy community. Engagement decreases immediately after self-disclosures of dissonance and dissonance experienced early on precedes departure from the community. Taken together, these results suggest that dissonance is followed by behavior change in the conspiracy community. In this light, we next discuss our empirical observation of how users manage their experiences of dissonance. We connect our findings to theoretical accounts of behavior change induced by dissonance.


\subsection{Cognitive Dissonance Reduction Strategies}
Our RQ3b results indicate that QAnon followers expressed dissonance about the legitimacy of the QAnon movement, unfulfilled expectations, ineffective practices and distrust in the heroes of the movement. What are the consequences of experiencing dissonance? Researchers posit that experiencing cognitive dissonance induces the state of psychological discomfort \cite{festinger1962theory}. To deal with this discomfort, people employ several dissonance reduction strategies \cite{festinger1962theory,mcgrath2017dealing}. For example, they trivialize the cause of dissonance and self-affirm their belief system \cite{simon1995trivialization}. Once an individual trivializes the point of contradiction, other discrepancies can no longer arouse dissonance. Consider this Reddit comment from our dataset for example: 

\begin{quote}
    \small \textit{I dont really care if Q is real. He has just reinforced my commitment to dig deeper. That's what a real Q dude would do no matter what. DIG DEEPER
}
\end{quote}

People also find strategies to rationalize the cause of dissonance \cite{mcgrath2017dealing}. 
Similarly, QAnon followers often rationalized Q's failed predictions by giving alternate explanations of the failures or creating more consonant interpretations of reality. 

\begin{quote}
    \small \textit{...[you] are not thinking critically in context of everything that comprises Q's message and content. He hasn't ``failed''. he's made statements that people have misinterpreted and then blamed on Q for the misinterpretations not being correct.}
\end{quote}


More importantly, however, researchers state that cognitive dissonance can lead to attitude and behavior changes rejecting previously held beliefs \cite{festinger1959cognitive}. Meaning, experiencing dissonance with conspiracies may lead people to abandon conspiracy beliefs and pave the way for recovery from conspiratorial worldview. 
Indeed, our analysis indicates that user contributions lowered after expressing dissonance, and that dissonance increased just before user's departure from the QAnon community. This is mirrored in the findings of our qualitative analysis. We found instances of users indicating that they were leaving the QAnon subreddit as a result of dissonance. 

\begin{quote}
    \small \textit{Q Anon is A psyop!!!! I am out, this board has been infiltrated. Something good, to something terrible, real quick. Don't fall for this Q stuff. Think for yourselves.
}
\end{quote}

\begin{quote}
    \small \textit{This is the last one for me I think. I got hyped for ``the memo'' I got hyped for ``raw footage''. Arrests of the cabal within the week or I am out.
}
\end{quote}

While acknowledging that dissonance may not always lead to positive behavior changes, our results suggest that exploring dissonance as a possible intervention for online conspiracy engagement is a promising future direction. Several other studies have explored dissonance based interventions \cite{freijy2013dissonance} for reducing implicit racial prejudice \cite{hing2002inducing} and promoting positive social behavior \cite{mckimmie2003m}. In the next section, we discuss how dissonance can be used as an intervention to motivate positive behavior change in conspiracy communities. 


\subsection{Intervention for Online Conspiracy Engagement}
In RQ3, we showed how users spontaneously disclosed experiences of dissonance and how this correlates with changes in behavior, especially focusing on the effects of disengaging from the conspiracy community, as demonstrated in RQ4.
Dissonance can also be introduced externally as an intervention to \emph{induce} such behavior. In fact, similar interventions based on ``hypocrisy paradigm'' \cite{fried1995hypocrisy}, that encourage participants to explore the difference between their internally held beliefs and their public expression, have been tested in settings ranging from mental health through addiction recovery to prejudice reduction \cite{hing2002inducing, freijy2013dissonance}. For example, participants with high implicit racial prejudice were asked to write an essay on racial justice and fairness. Publicly expressing views contradictory to the implicit beliefs led the participants to reduce the prejudicial behavior \cite{hing2002inducing}. 
Our results in RQ3b show that such ``hypocrisy'' exists in the QAnon conspiracy community. For example, the scenario where users want to be part of the QAnon community while at the same time dislike some aspects of the QAnon movement or doubt the QAnon heroes. 


The methods in RQ1 and RQ2 offer ways to systematically compile social imaginaries from the point of view of the conspiracist themselves. This may help design community centered hypocrisy interventions based strategies that could nudge the conspiracists to explore the differences between the social imaginaries of the community and their internally held beliefs. For example, building on the qualitative analysis of fracture points in RQ3, one could build interventions that question the infallibility of heroes and the promises made by the movement leaders. Moreover, our results may also help contextualize past successful interventions within the social imaginaries of a specific community and to recast them as hypocrisy interventions, such as questioning the efficacy of the movement to rigorously derive truth \mbox{\cite{Cook2018}} or to be effective against foes \mbox{\cite{Stojanov2020}}.


Furthermore, 
the computational framework in RQ3 can be used to identify the central causes of dissonance in the community. These fracture points can be insisted or expanded upon in strategic interventions. Our results can also inform which interventions might not be successful in the community. For example, in QAnon, ``MSM'' (mainstream media) is heavily distrusted and is considered as a foe. Interventions citing news articles from mainstream sources maybe met with instant criticism, despite of their credibility. Finally, RQ4 offers ways to measure the outcomes of interventions, and therefore to select ones that are most effective. Hence, conspiracy social imaginaries and computational dissonance detection offer powerful tools to design contextually-informed interventions.

\subsubsection{\textbf{Ethical Considerations}} However, researchers need to consider social, psychological and ethical consequences of designing such intervention systems. For example, \citeauthor{sunstein2009conspiracy} argue that sowing the seeds of doubt in conspiracy theory communities is most (or perhaps only) effective when done from within the community \cite{sunstein2009conspiracy}. Skepticism coming from outsiders may be deemed illegitimate, or even be construed as part of a larger conspiracy attempting to undermine the conspiracists' truth \cite{Lewandowsky2013}. Dissonance causes psychological discomfort, and therefore interventions inducing dissonance should weigh harms against benefits. It is also important to consider \emph{whether} certain conspiracies need intervention by accounting for the researchers' socio-political biases. These are but few of the ethical issues that intervention designers should consider before interacting directly with social media participants.


In sum, designing interventions for conspiracy communities is a complex socio-technical and even, a political problem that needs careful considerations of research ethics. 
Specifically, the design of hypocrisy based interventions described in Section 8.2 will require multi-disciplinary effort of computer scientists, social psychologists and ethics and privacy experts who could anticipate various technological, social and privacy implications of the interventions. A careful design of interventions will involve deriving least invasive intervention strategies and systematically evaluating the intervention impact while monitoring for unintended consequences. We do not propose dissonance as an ultimate, or only solution for recovery from conspiracy theories but rather hope to initiate a dialogue about carefully designed interventions supported by our findings.

\subsection{Conspiracy Belief as Collective Intelligence}



Our findings in RQ3 and RQ4 enrich previous quantitative work on online conspiracy communities. Users who join conspiracy communities do so through contacts with existing members in conspiracy-neutral spaces \cite{phadke2021makes}. On the one hand, there is empirical evidence of both self-selection of users into the conspiracy community and of shunning users from non-conspiracy communities. On the other hand, users who remain for a sufficiently long time may become overly committed \cite{samory2018conspiracies}. In RQ4 we show that there exists a critical time between users joining the community and committing to it in the long term, where arbitration in terms of self-disclosures and feedback received from the members affects user participation. Especially, users who exhibit heightened experiences of dissonance early on in their tenure in the community tend to leave, whereas users who remain may have successfully resolved existential conflicts between their beliefs and those of the community. 

Hence, we suggest moving towards a view of conspiracy theorizing as a collaborative pursuit between individuals and communities. On the one hand, interventions should take into account both individual and collective beliefs, as well as the possible stages of the relationships between users and communities. On the other hand, the study of conspiracy theories may draw insights from research on arbitration between individuals and communities, such as community migration, inter-community conflict, and moderation, to unpack the relationship between adopting conspiracy beliefs and joining conspiracy theory communities.


\subsection{Implications for Studying Conspiracy Semiotics}
Previous computational research studying language of conspiracy discussions mostly focuses on uncovering overarching narrative and argumentative elements in discourse \cite{samory2018government}. Our methods can complement and enrich these techniques by revealing semiotic patterns in conspiracy discussions. 


In our work, we propose a novel approach to uncover symbolic phrases that share common meanings inside the conspiracy communities. By combining manual verification with computationally creating dynamic representation of phrases, we automate the snowballing technique for uncovering symbolic language. Further by characterizing the insider language correlates across the dimensions of QAnon social imaginaries, we also provide a way for interpreting underlying meanings. Our methods and the lexicon QAnon Canon---the lexicon encoding semiotics in QAnon---together provide a toolbox for social-psychology and CSCW researchers to investigate language and expressions in QAnon like conspiracy communities within and across social media platforms. Although QAnon Canon is built from Reddit discussions, it can be useful in studying QAnon communities on other platforms as well, given that the general QAnon discourse is inspired by the same social imaginaries proposed in Q-drops. Specifically given the recent riots on U.S Capitol carried out by QAnon followers, our lexicon can be used to identify the dimensions of social imaginaries that were most prominent in mobilizing the collective action. Using the techniques proposed in this work, social computing researchers can further expand the lexicon to incorporate semiotics in other conspiracy theories.


\section{Limitations and Future Directions}
Our work has some limitations that may offer directions for future research. First, the time frame of our study is constrained between the start of QAnon in 2017 to the ban of the QAnon subreddits in 2018. Since then, the QAnon movement has grown across various platforms such as Gab, Voat, and Facebook. Though our observations are temporally constrained, our analytical framework is general and could be applied to study dissonance self-disclosures across those different spaces and time frames. A further limitation concerns the procedure of extracting coded language from Q-drops. We relied on variety of previous research and online sources to understand the social imaginaries surrounding such language. 
While this allows us to be confident about the phrases captured in the seed lexicon, it is possible that we overlooked some other coded words due to lack of context. Though time consuming, a potentially more comprehensive process could iteratively surface coded words from Q-drops and interpret them through the online discussion comments that reference them. Moreover, while our classifier gives precision above 0.7 in a complex three class classification problem, it relies on relatively simple features. Its performance would likely improve further after incorporating stylistic and meta-linguistic features and a larger labeled dataset. Finally, while our RQ4 analysis provides initial evidence about changes in user engagement following dissonance disclosures, we do not make causal claims. Studies aiming at deriving causal connections between disclosures of dissonance and behavior change should adopt controlled experimental designs and should account for potential confounders.

\section{Conclusion}
In this work, we studied the social imaginaries within the QAnon conspiracy theory community. We uncovered five dimensions of QAnon social imaginaries and created a lexicon capturing coded language across the five dimensions. We used this lexicon to identify self-disclosures of cognitive dissonance inside the QAnon community and typified dissonant expressions along the social imaginaries. We further provided evidence that user contributions inside QAnon communities decrease immediately after self-disclosures of dissonance, and that high levels of experienced dissonance correlate with users ultimately leaving the communities. Our results show that users \emph{do} express dissonance inside their communities and dissonance can be explored further as possible intervention for online conspiracy engagement.  

\section{Acknowledgments}
We want to acknowledge the valuable feedback from the members of Social Computing Lab at Virginia Tech and University of Washington, Seattle towards strengthening this work. This project was partially funded by the Minerva Research Initiative.

\bibliographystyle{ACM-Reference-Format}
\bibliography{sample-base}

\appendix

\section{Appendix}

\subsection{RQ3a Codebook used for Manually Labeling Samples}
\label{sec:codebook}

Our codebook was developed over multiple pilot labeling experiments where the authors went back and forth referring to Q-drops and Reddit comments to understand discourse in the QAnon community. 
\begin{enumerate}
    \item \textbf{Belief: } It is belief when atleast part of the message shows support for Q/the movement/the subreddit/the social imaginaries or shows interest in the QAnon activities or shows strong opposition/dislike of the enemies of QAnon. It is not belief when the message presents belief in something outside of QAnon (aliens, religion). 
    \item \textbf{Dissonance: } It is dissonance if at least part of the message is such that the speaker contradicts, doubts or expresses uncertainty about the social imaginaries or argues against the movement/redditors/a specific redditor when the latter defends the social imaginaries or proposes alternatives as being preferable to Q (god, other celebrities, politicians). It is not dissonance when the speaker confronts individual users without reference to the social imaginaries or argues for the relative/speculative nature of the social imaginaries, without denying its usefulness or truth. 
    \item \textbf{Neutral: } The comment is neutral if the message does not contain any reference to beliefs/dissonance with respect to QAnon. 
\end{enumerate}

While labeling, we also skipped a few samples that we could not understand or interpret with confidence.


\subsection{RQ3a Classification Additional Reports}

\subsubsection{Experiments with various Prediction Pooling Strategies}
\label{sec:pooling_strats}
While we select consensus pooling for the identifying dissonance, we also experimented with other pooling strategies described below. We also report the precision and recall results on validation set for each of the strategies in Table \ref{tab:pooling_expt}. 

\begin{enumerate}
    \item \textbf{Max Pooling: } Consider the prediction with highest probability value of both classifiers
    \item \textbf{Average Pooling: } Calculate the average of class probabilities generated by both classifiers and consider the class with maximum averaged probability
    \item \textbf{Stacking: } Train a third classifier based on both, the predictions of two classifiers and all the features. Consider the predictions of the third classifier as final
\end{enumerate}

\begin{table*}[]
\centering
\resizebox{0.7\textwidth}{!}{%
\begin{tabular}{@{}ccc|cc|cc@{}}
\toprule
          & \multicolumn{2}{c|}{\textbf{RF+XGB Max Pooling}} & \multicolumn{2}{c|}{\textbf{RF+XGB Avg Pooling}} & \multicolumn{2}{c}{\textbf{RF+XGB Stacking}} \\ \midrule
\textbf{} & \multicolumn{2}{c|}{validation}                  & \multicolumn{2}{c|}{validation}                  & \multicolumn{2}{c}{validation}               \\
\textbf{}           & precision & recall & precision & recall & precision & recall \\
\textbf{belief}     & 0.63      & 0.59   & 0.62      & 0.61   & 0.61      & 0.63   \\
\textbf{dissonance} & 0.61      & 0.62   & 0.59      & 0.61   & 0.65      & 0.64   \\
\textbf{neutral}    & 0.67      & 0.70   & 0.69      & 0.68   & 0.75      & 0.72  
\end{tabular}%
}
\caption{Table showing validation performances with different prediction pooling strategies. }
\label{tab:pooling_expt}
\end{table*}

\subsubsection{Stability of Validation Performance}
\begin{figure*}[thb]
    \centering 
\begin{subfigure}{0.30\textwidth}
  \includegraphics[width=\linewidth]{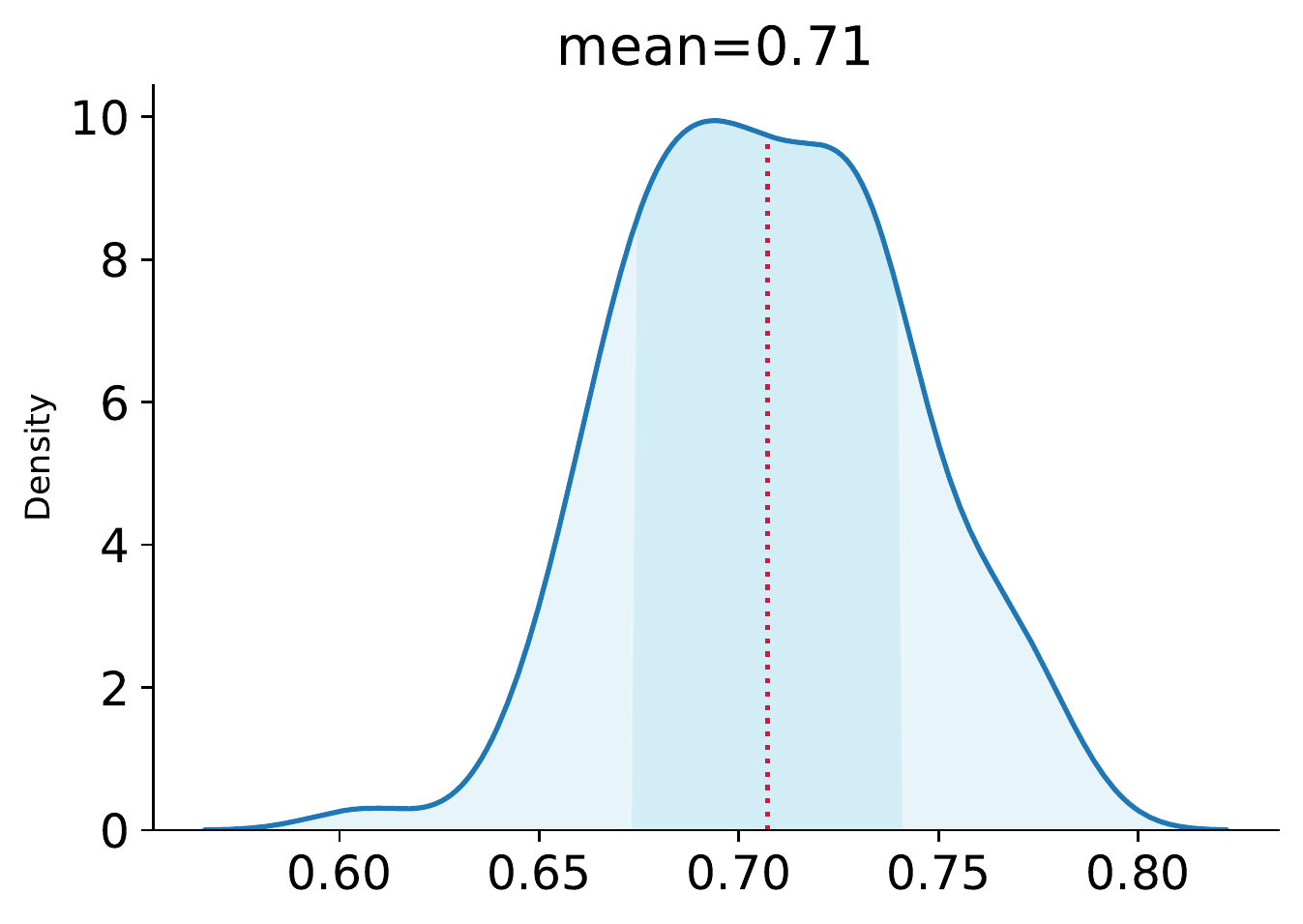}
  \caption{Belief Precision}
  \label{fig:bp}
\end{subfigure}\hfil 
\begin{subfigure}{0.30\textwidth}
  \includegraphics[width=\linewidth]{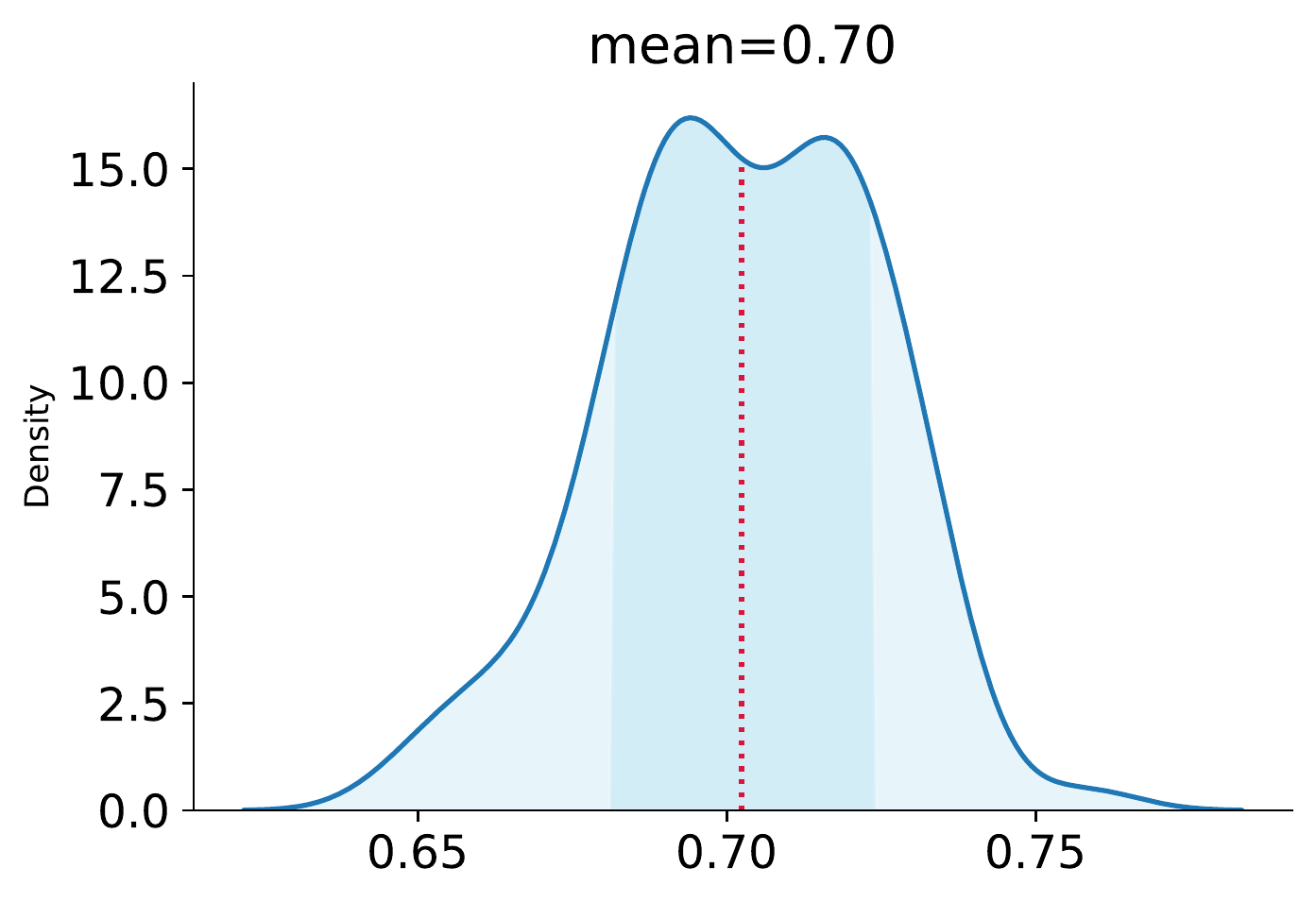}
  \caption{Dissonance Precision}
  \label{fig:dp}
\end{subfigure}\hfil 
\begin{subfigure}{0.30\textwidth}
  \includegraphics[width=\linewidth]{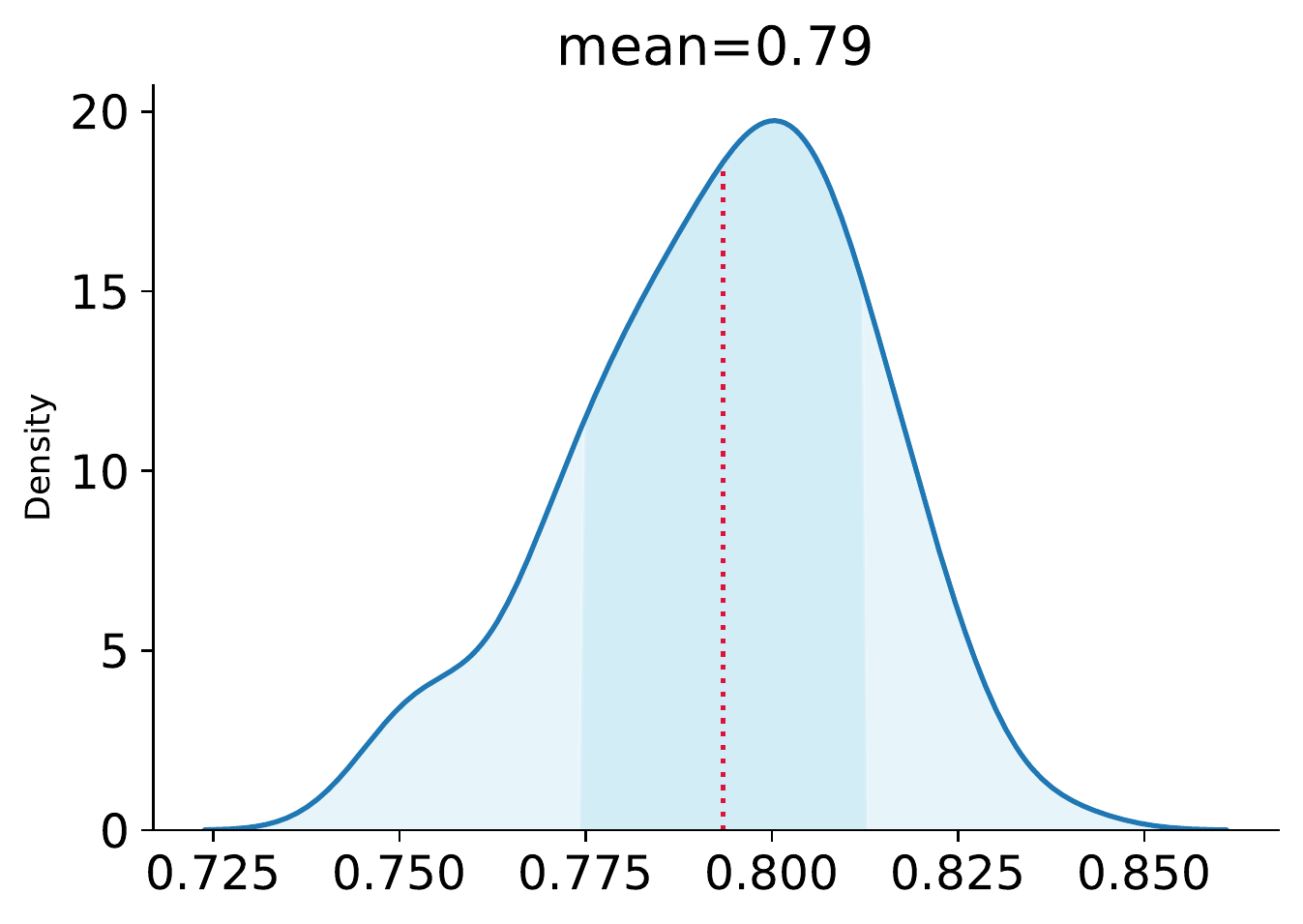}
  \caption{Neutral Precision}
  \label{fig:ip}
\end{subfigure}

\medskip
\begin{subfigure}{0.30\textwidth}
  \includegraphics[width=\linewidth]{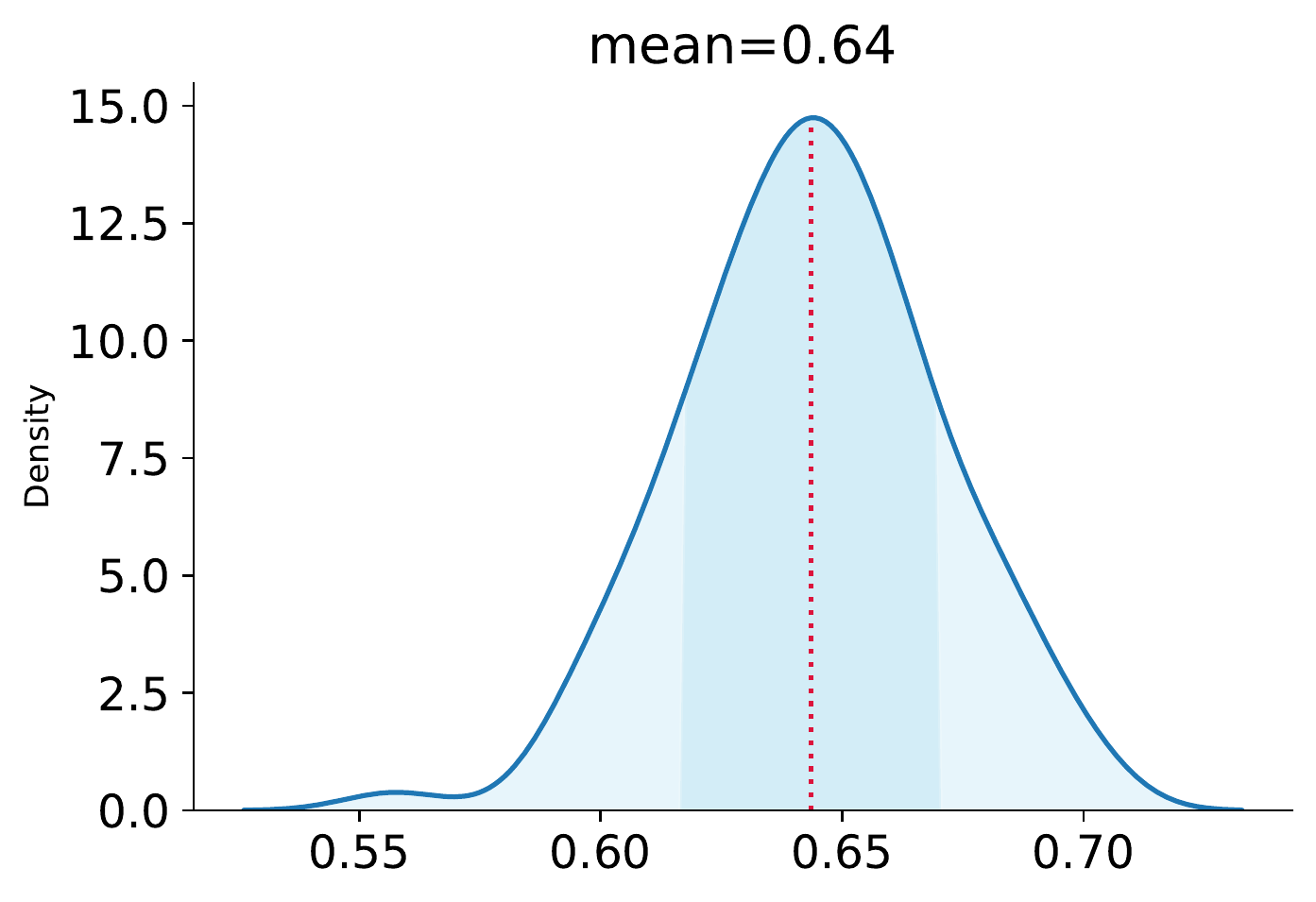}
  \caption{Belief Recall}
  \label{fig:br}
\end{subfigure}\hfil 
\begin{subfigure}{0.30\textwidth}
  \includegraphics[width=\linewidth]{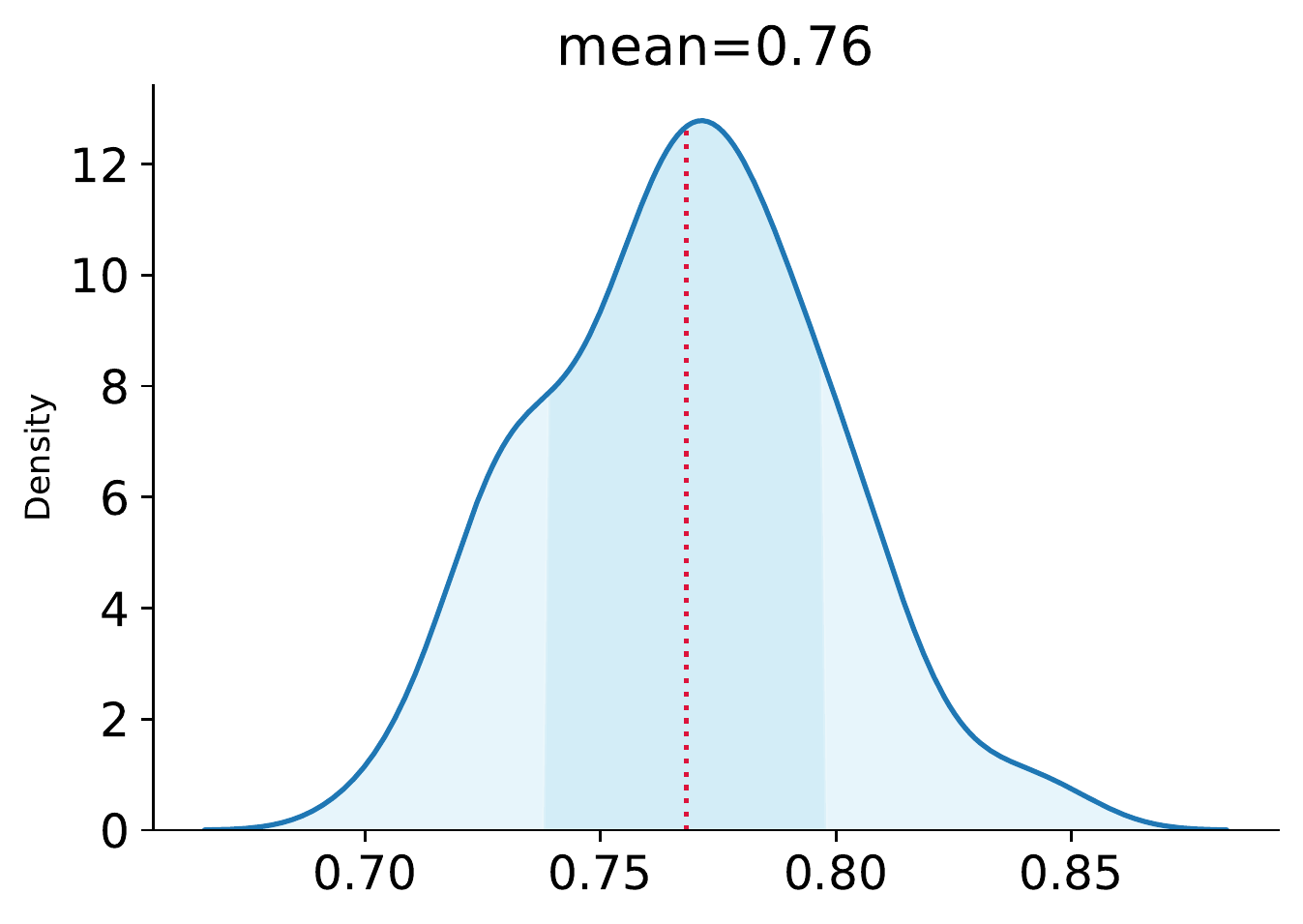}
  \caption{Dissonance Recall}
  \label{fig:dr}
\end{subfigure}\hfil 
\begin{subfigure}{0.30\textwidth}
  \includegraphics[width=\linewidth]{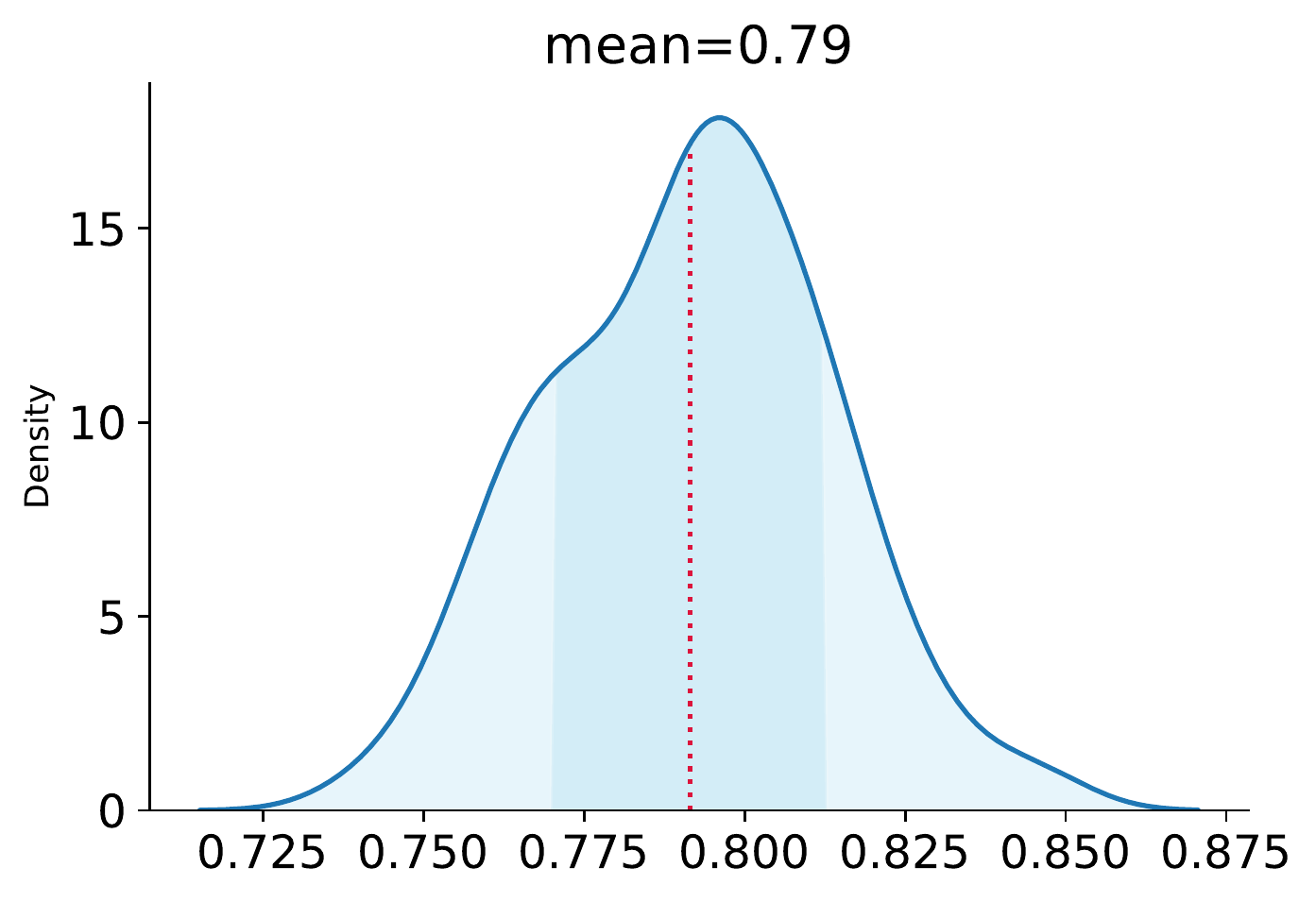}
  \caption{Neutral Recall}
  \label{fig:ir}
\end{subfigure}
\caption{Validation precision and recall distributions of three classes over 100 training iterations. The dotted red line represents the mean and the shaded area around it represents the standard deviation.}
\label{fig:validation}
\end{figure*}

We tested the validation scores for final consensus classifier over 100 training iterations. Training tree based classifiers involves certain degree of randomness. Hence, we calculate the average precision and recall scores for all classes over 100 training iterations. Figure \ref{fig:validation} displays the precision recall values for all classes over 100 training iterations. Overall, the plots suggests our validation scores are balanced and thus our model has less chances of overfitting. 

\subsection{RQ4 User Engagement Additional Reports}

\subsubsection{ITS Analysis of Dissonance in Predicted Data}
\label{sec:predicted_its}

\begin{figure*}[t]
    \centering
    \includegraphics[width=0.9\textwidth]{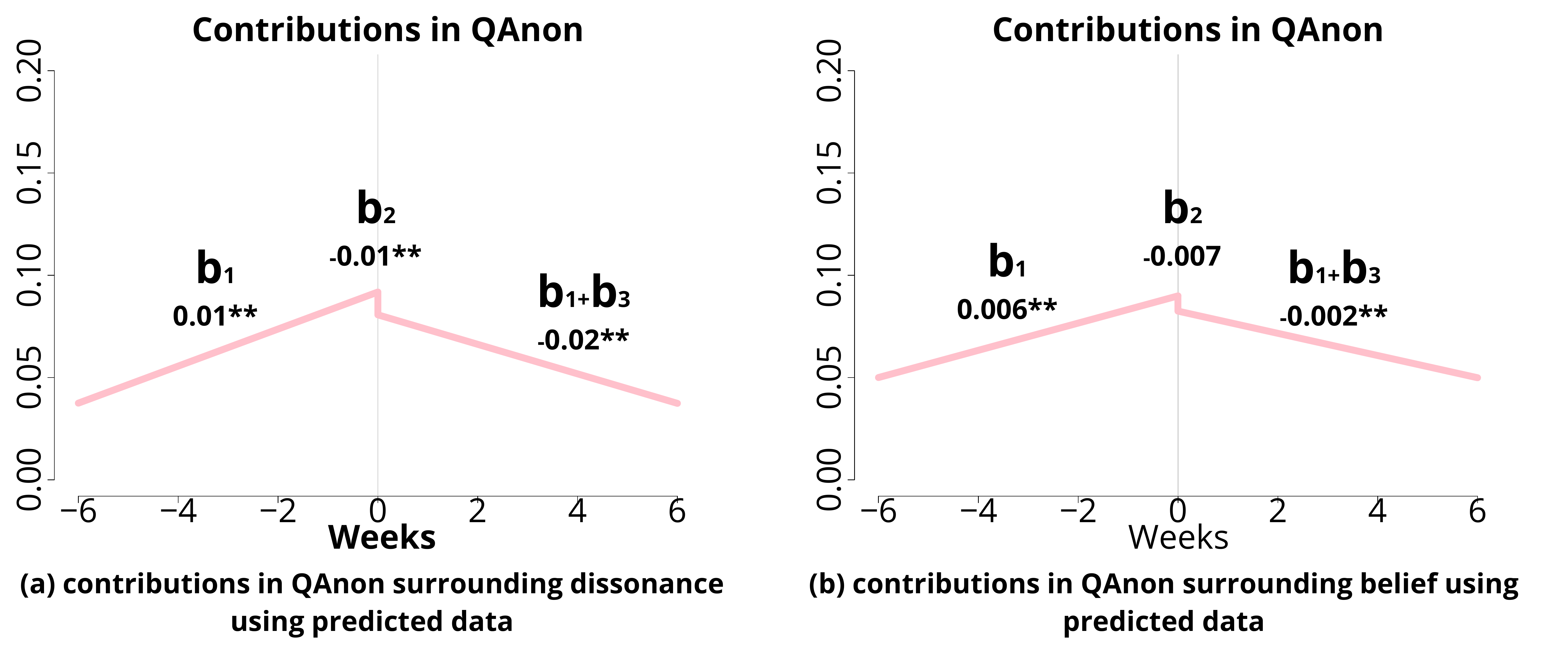}
    \caption{ITS based on automatically labeled comments. We observe similar trends as manually labeled comments reported in Figure \ref{fig:its_labeled}.}
    \label{fig:its_predicted}
\end{figure*}

While in RQ4 we present the ITS analysis of the labeled data, we also repeat the same experiment for the entire set of predicted comment. Figure \ref{fig:its_predicted} displays trends plotted using all predicted instances of dissonance and belief. The trends in the predicted data follow the same pattern as the results presented in Figure \ref{fig:its_labeled} in the main paper. 

\subsubsection{Regression Analysis of User Engagement }
Here we present the results of regression analyses that further corroborate our observations of decreased engagement in the ITS analysis. We predict the number of remaining days that the user will spend on the subreddit (Table \ref{tab:nbdeath}), number of comments user will likely make before exiting the community (Table \ref{tab:olsdeath}) and the binary outcome of whether the user will remain in the community for more than 10 days (Table \ref{tab:logitdeath}). 

\begin{table}[] 

\centering
\footnotesize
\begin{tabular}{lclc}
\toprule
\textbf{Dep. Variable:} &        days until user leaves         & \textbf{  No. Observations:  } &     1773    \\
\textbf{Model:}         & NegativeBinomial & \textbf{  Df Residuals:      } &     1767    \\
\textbf{Method:}        &       MLE        & \textbf{  Df Model:          } &        5    \\
\textbf{Date:}          & Wed, 14 Apr 2021 & \textbf{  Pseudo R-squ.:     } &  0.09317    \\
\textbf{Time:}          &     10:52:21     & \textbf{  Log-Likelihood:    } &   -7910.1   \\
\textbf{converged:}     &       True       & \textbf{  LL-Null:           } &   -8722.7   \\
\bottomrule
\end{tabular}
\begin{tabular}{lcccccc}
                    & \textbf{coef} & \textbf{std err} & \textbf{z} & \textbf{P$> |$z$|$} & \textbf{[0.025} & \textbf{0.975]}  \\
\midrule
\textbf{const}      &       3.6545  &        0.014     &   261.392  &         0.000        &        3.627    &        3.682     \\
\textbf{born}       &      -0.0648  &        0.018     &    -3.686  &         0.000        &       -0.099    &       -0.030     \\
\textbf{max(score)}   &       0.0391  &        0.015     &     2.645  &         0.008        &        0.010    &        0.068     \\
\textbf{dissonance} &      -0.0301  &        0.014     &    -2.141  &         0.032        &       -0.058    &       -0.003     \\
\textbf{belief}     &      -0.0090  &        0.014     &    -0.642  &         0.521        &       -0.037    &        0.019     \\
\textbf{created}    &      -0.7951  &        0.019     &   -41.482  &         0.000        &       -0.833    &       -0.758     \\
\textbf{alpha}      &       0.3141  &        0.013     &    24.610  &         0.000        &        0.289    &        0.339     \\
\bottomrule
\end{tabular}
\caption{Negative binomial regression results, predicting in how many days the user will leave the community. The maximum comment score is positively correlated with longer permanence on the subreddit. Disclosures of dissonance, instead, are negatively correlated. Hence, disclosures of dissonance signal that users will leave the subreddit soon. } 
    \label{tab:nbdeath}
\end{table}

\begin{table}[]
\centering
\footnotesize
\begin{tabular}{lclc}
\toprule
\textbf{Dep. Variable:}     & log(comments left) & \textbf{  R-squared:         } &     0.507   \\
\textbf{Model:}             &        OLS         & \textbf{  Adj. R-squared:    } &     0.504   \\
\textbf{Method:}            &   Least Squares    & \textbf{  F-statistic:       } &     226.5   \\
\textbf{Date:}              &  Wed, 14 Apr 2021  & \textbf{  Prob (F-statistic):} & 3.15e-264   \\
\textbf{Time:}              &      11:56:51      & \textbf{  Log-Likelihood:    } &   -2531.5   \\
\textbf{No. Observations:}  &         1773       & \textbf{  AIC:               } &     5081.   \\
\textbf{Df Residuals:}      &         1764       & \textbf{  BIC:               } &     5130.   \\
\textbf{Df Model:}          &            8       & \textbf{                     } &             \\
\bottomrule
\end{tabular}
\begin{tabular}{lcccccc}
                            & \textbf{coef} & \textbf{std err} & \textbf{t} & \textbf{P$> |$t$|$} & \textbf{[0.025} & \textbf{0.975]}  \\
\midrule
\textbf{const}              &       4.6015  &        0.024     &   191.560  &         0.000        &        4.554    &        4.649     \\
\textbf{born}               &       0.4513  &        0.029     &    15.552  &         0.000        &        0.394    &        0.508     \\
\textbf{min(score)}           &       0.0608  &        0.024     &     2.519  &         0.012        &        0.013    &        0.108     \\
\textbf{max(score)}           &       0.0608  &        0.024     &     2.484  &         0.013        &        0.013    &        0.109     \\
\textbf{avg($\mathcal{D}$)} &      -0.0185  &        0.059     &    -0.311  &         0.756        &       -0.135    &        0.098     \\
\textbf{max($\mathcal{D}$)} &       0.0339  &        0.052     &     0.655  &         0.513        &       -0.068    &        0.135     \\
\textbf{dissonance}         &      -0.0691  &        0.033     &    -2.095  &         0.036        &       -0.134    &       -0.004     \\
\textbf{belief}             &       0.0056  &        0.028     &     0.198  &         0.843        &       -0.050    &        0.061     \\
\textbf{created}            &      -1.2059  &        0.029     &   -41.294  &         0.000        &       -1.263    &       -1.149     \\
\bottomrule
\end{tabular}
\begin{tabular}{lclc}
\textbf{Omnibus:}       & 60.154 & \textbf{  Durbin-Watson:     } &    1.950  \\
\textbf{Prob(Omnibus):} &  0.000 & \textbf{  Jarque-Bera (JB):  } &   81.072  \\
\textbf{Skew:}          & -0.354 & \textbf{  Prob(JB):          } & 2.49e-18  \\
\textbf{Kurtosis:}      &  3.771 & \textbf{  Cond. No.          } &     4.99  \\
\bottomrule
\end{tabular}
\caption{Ordinary least squares regression results, predicting the log-scaled number of comments that the user will post in the future. Similarly to the negative binomial regression, comment scores are positively correlated with more future comments, while disclosures of dissonance correlate negatively. } 
    \label{tab:olsdeath}

\end{table}

\begin{table}[]
\centering
\footnotesize
\begin{tabular}{lclc}
\toprule
\textbf{Dep. Variable:}     &        remains after 10 days         & \textbf{  No. Observations:  } &     1773    \\
\textbf{Model:}             &      Logit       & \textbf{  Df Residuals:      } &     1766    \\
\textbf{Method:}            &       MLE        & \textbf{  Df Model:          } &        6    \\
\textbf{Date:}              & Wed, 14 Apr 2021 & \textbf{  Pseudo R-squ.:     } &   0.3708    \\
\textbf{Time:}              &     11:03:12     & \textbf{  Log-Likelihood:    } &   -508.19   \\
\textbf{converged:}         &       True       & \textbf{  LL-Null:           } &   -807.64   \\
\bottomrule
\end{tabular}
\begin{tabular}{lcccccc}
                            & \textbf{coef} & \textbf{std err} & \textbf{z} & \textbf{P$> |$z$|$} & \textbf{[0.025} & \textbf{0.975]}  \\
\midrule
\textbf{const}              &       3.1499  &        0.170     &    18.528  &         0.000        &        2.817    &        3.483     \\
\textbf{born}               &      -0.2703  &        0.080     &    -3.388  &         0.001        &       -0.427    &       -0.114     \\
\textbf{avg(score)}           &       0.1447  &        0.075     &     1.938  &         0.053        &       -0.002    &        0.291     \\
\textbf{min(score)}           &       0.1694  &        0.070     &     2.424  &         0.015        &        0.032    &        0.306     \\
\textbf{avg($\mathcal{D}$)} &       0.3087  &        0.173     &     1.786  &         0.074        &       -0.030    &        0.648     \\
\textbf{max($\mathcal{D}$)} &      -0.3363  &        0.169     &    -1.993  &         0.046        &       -0.667    &       -0.006     \\
\textbf{created}            &      -2.4774  &        0.171     &   -14.523  &         0.000        &       -2.812    &       -2.143     \\
\bottomrule
\end{tabular}
\caption{Logistic regression results, predicting whether the user remains in the community for more than 10 days after posting the 100th comment. The results corroborate the previous analyses. In particular, the negative coefficient for the maximum dissonance index $\mathcal{D}$ indicates that users who experience the highest dissonance are the ones who leave the community the soonest.} \label{tab:logitdeath}
\end{table}

\end{document}